\documentclass{article}[12pt] 
\usepackage{ amsmath, amsthm, amssymb } 
\usepackage{color} 
\usepackage{cite}

\newcommand{\be} { \begin{equation} } 
\newcommand{\ee} { \end{equation} } 

\newcommand{\labbel}[1] { \label{#1} } 

\newcommand{\D}{\mathfrak{D}}
\newcommand{\I}{\mathcal{I}}
\newcommand{\J}{\mathcal{J}}

\newcommand{\N}{\mathbb{N}}

\newcommand{\Sunriseone}[1]{ 
\mbox{\parbox{2.5cm}{\hspace{0.25cm} 
\begin{picture}(2,1) 
\thicklines 
\put(0.3,0.5){\vector(1,0){0.1}} 
\put(0.5,0.5){\line(1,0){1}} 
\put(0,0.5){\line(1,0){0.5}} 
\put(1.5,0.5){\line(1,0){0.5}} 
\put(1,0.5){\circle{2}} 
\put(0.85,0.08){$m$} 
\put(0.25,0.7){\makebox(0,0)[b]{$#1$}} 
\end{picture} 
}} 
\hfill} 

\newcommand{\Sunrisetwo}[1]{ 
\mbox{\parbox{2.5cm}{\hspace{0.25cm} 
\begin{picture}(2,1) 
\thicklines 
\put(0.3,0.5){\vector(1,0){0.1}} 
\put(0.5,0.5){\line(1,0){1}} 
\put(0,0.5){\line(1,0){0.5}} 
\put(1.5,0.5){\line(1,0){0.5}} 
\put(1,0.5){\circle{2}} 
\put(0.85,1.12){$m_1$} 
\put(0.85,0.60){$m_2$} 
\put(0.85,0.08){$m_3$} 
\put(0.25,0.7){\makebox(0,0)[b]{$#1$}} 
\end{picture} 
}} 
\hfill}

\newcommand{\threeban}[1]{ 
\mbox{\parbox{2.5cm}{\hspace{0.25cm} 
\begin{picture}(2,1) 
\thicklines 
\put(0.3,0.5){\vector(1,0){0.1}} 
\put(0,0.5){\line(1,0){0.45}} 
\put(1.55,0.5){\line(1,0){0.45}} 
\put(1,0.5){\circle{2}} 
\qbezier(0.45, 0.5)(1, 1)(1.55, 0.5)
\qbezier(0.45, 0.5)(1, 0.05)(1.55, 0.5)
\put(0.25,0.7){\makebox(0,0)[b]{$#1$}} 
\end{picture} 
}} 
\hfill}

\newcommand{\triangleThree}[3]{
\mbox{\parbox{3cm}{\hspace{-0.1cm}
\begin{picture}(2.5,1.4)
\thicklines
\put(0,0.7){\line(1,0){0.5}}
\put(1.0,1.2){\line(1,0){0.7}}
\put(1.0,0.2){\line(1,0){0.7}}
\thinlines
\qbezier(1.0, 1.2)(1.5, 0.75)(1.0, 0.2)
\put(1.0,0.2){\line(0,1){1.0}}
\put(0.3,0.7){\vector(1,0){0.1}}
\put(1.5,0.2){\vector(1,0){0.1}}
\put(1.5,1.2){\vector(1,0){0.1}}
\put(0.5,0.7){\line(1,1){0.5}}
\put(0.5,0.7){\line(1,-1){0.5}}
\put(1,1.2){\line(1,0){0.5}}
\put(1,0.2){\line(1,0){0.5}}
\put(0.25,0.9){\makebox(0,0)[b]{$#1$}}
\put(1.85,1.2){\makebox(0,0)[l]{$#2$}}
\put(1.85,0.2){\makebox(0,0)[l]{$#3$}}
\end{picture}
}}
\hfill}

\newcommand{\triangleMass}[3]{
\mbox{\parbox{3cm}{\hspace{-0.1cm}
\begin{picture}(2.5,1.4)
\thicklines
\put(0,0.7){\line(1,0){0.5}}
\put(1.0,1.2){\line(1,0){0.7}}
\put(1.0,0.2){\line(1,0){0.7}}
\put(1.0,0.2){\line(0,1){1.0}}
\put(0.3,0.7){\vector(1,0){0.1}}
\put(1.5,0.2){\vector(1,0){0.1}}
\put(1.5,1.2){\vector(1,0){0.1}}
\put(0.5,0.7){\line(1,1){0.5}}
\put(0.5,0.7){\line(1,-1){0.5}}
\put(1,1.2){\line(1,0){0.5}}
\put(1,0.2){\line(1,0){0.5}}
\thinlines
\qbezier(1.0, 1.2)(0.5, 1.2)(0.5, 0.7)
\put(0.25,0.9){\makebox(0,0)[b]{$#1$}}
\put(1.85,1.2){\makebox(0,0)[l]{$#2$}}
\put(1.85,0.2){\makebox(0,0)[l]{$#3$}}
\end{picture}
}}
\hfill}

\newcommand{\trianglecross}[3]{
\mbox{\parbox{3cm}{\hspace{-0.1cm}
\begin{picture}(2.5,1.4)
\thicklines
\put(1.45,0.82){$m$}
\put(1.45,0.42){$m$} 
\put(1.5,1.2){\line(-1,-2){0.5}}
\put(0,0.7){\line(1,0){0.5}}
\put(1.0,1.2){\line(1,-2){0.18}}
\put(1.5,0.2){\line(-1,2){0.18}}
\thinlines
\put(0.3,0.7){\vector(1,0){0.1}}
\put(1.8,0.2){\vector(1,0){0.1}}
\put(1.8,1.2){\vector(1,0){0.1}}
\put(0.5,0.7){\line(1,1){0.5}}
\put(0.5,0.7){\line(1,-1){0.5}}
\put(1,1.2){\line(1,0){0.5}}
\put(1,0.2){\line(1,0){0.5}}
\put(1.5,1.2){\line(1,0){0.5}}
\put(1.5,0.2){\line(1,0){0.5}}
\put(0.25,0.9){\makebox(0,0)[b]{$#1$}}
\put(2.05,1.2){\makebox(0,0)[l]{$#2$}}
\put(2.05,0.2){\makebox(0,0)[l]{$#3$}}
\end{picture}
}}
\hfill}

\begin{document} 
\setlength{\unitlength}{1.3cm} 
\begin{titlepage}
\vspace*{-1cm}
\begin{flushright}
TTP15-034
\end{flushright}                                
\vskip 3.5cm
\begin{center}
\boldmath
{\Large\bf Integration by parts identities in integer numbers of dimensions. \\ 
A criterion for decoupling systems of differential equations\\[3mm] }
\unboldmath
\vskip 1.cm
{\large Lorenzo Tancredi}$^{b,}$
\footnote{{\tt e-mail: lorenzo.tancredi@kit.edu}} 
\vskip .7cm
{\it $^b$ Institute for Theoretical Particle Physics, KIT, Karlsruhe, Germany } 
\end{center}
\vskip 2.6cm

\begin{abstract}
Integration by parts identities (IBPs) can be used to express large numbers of apparently different
$d$-dimensional Feynman Integrals in terms of a small subset of so-called master integrals (MIs). 
Using the IBPs one can moreover show that the MIs fulfil linear systems of coupled differential equations
in the external invariants. With the increase in number of loops and external legs,
one is left in general with an increasing number of MIs and consequently also with an increasing number 
of coupled differential equations, 
which can turn out to be very difficult to solve. 
In this paper we show how studying the IBPs in fixed integer numbers of dimension $d=n$ with $n \in \N$
one can extract the information useful to determine a new basis of MIs, whose differential equations
decouple as $d \to n$ and can therefore be more easily solved as Laurent expansion in $(d-n)$.

\end{abstract}
\vfill
\end{titlepage}                                                                
\newpage

\section {Introduction } \labbel{sec:intro} \setcounter{equation}{0} 
\numberwithin{equation}{section}
Dimensionally regularised~\cite{'tHooft:1972fi,Cicuta:1972jf,Bollini:1972ui} 
Feynman integrals fulfil different identities, among which the most
notable ones are the so-called integration by parts identities 
(IBPs)~\cite{Tkachov:1981wb, Chetyrkin:1981qh}.
Given a family of Feynman integrals, the IBPs can be used to write down a large system of linear equations
with rational coefficients, which contain the Feynman integrals of that family as unknowns. 
By solving algebraically the system a large number of apparently
different Feynman integrals can be expressed 
in terms of a much smaller basis of independent integrals dubbed master integrals (MIs).
In realistic applications the number of such equations can grow very fast, requiring the use of computer
algebra in order to handle the complexity of the resulting expressions.
There are different public and private implementations which allow to perform
the reduction to MIs in a completely automated way~\cite{Anastasiou:2004vj,Smirnov:2008iw,
Studerus:2009ye,vonManteuffel:2012np} based on the so-called Laporta algorithm~\cite{Laporta:1996mq,Laporta:2001dd}.

The IBPs can be used to prove that dimensionally regularised Feynman integrals
fulfil linear systems of first order differential equations 
in the external 
invariants~\cite{Kotikov:1990kg,Bern:1993kr,Remiddi:1997ny,Caffo:1998du,Gehrmann:1999as}.
A thorough review of the method can be found in~\cite{Argeri:2007up}.
Considering a Feynman graph which is reduced to $N$ independent MIs, by direct
use of the IBPs one can
derive a system of $N$ coupled linear first order differential equations for the latter,
which can be rephrased as an $N$-th order differential equation for any of the MIs. 
Supplemented with $N$ independent boundary conditions, the system of differential
equations contains all information needed for numerical or analytical calculations
of the MIs.
Indeed, in the general case, the analytical solution of an $N$-th order differential equation 
is a very non-trivial mathematical problem.

It has been observed that, in many cases of practical interest, 
a substantial simplification of the problem occurs when studying
the behaviour of the system of differential equations 
as the space-time dimension parameter $d$ approaches $4$, which is also the physically 
relevant case. Usually we are indeed not interested in an exact solution
for the MIs as functions of $d$, but instead in the coefficients of their Laurent expansion for $d \approx 4$.
In~\cite{Gehrmann:2000zt,Gehrmann:2001ck}, and in many subsequent applications of
the differential equation method, it was shown that it is often possible to choose a basis of
MIs such that the differential equations take a simpler triangular form
in the limit $d \to 4$. If this is possible, the problem of integrating the system of
differential equations simplifies substantially, reducing \textsl{de facto}, at every order in $(d-4)$, to 
$N$ subsequent integrations by quadrature. 
Experience showed that, whenever such a form is achievable, the differential
equations can be integrated in terms of a particular class of special functions,
the multiple polylogarithms (MPLs)~\cite{Goncharov,Remiddi:1999ew,Gehrmann:2000zt}.
The latter have been studied extensively by both mathematicians and physicists and
routines for their fast and precise numerical evaluation are 
available since some time~\cite{Gehrmann:2001pz,Gehrmann:2001jv,Vollinga:2004sn}.
Disclosing their algebraic properties allowed furthermore the development of very powerful
tools for the analytical manipulations of these functions~\cite{Duhr:2011zq,Duhr:2012fh,Panzer:2014caa}.

More recently it has been shown~\cite{Kotikov:2010gf,Henn:2013pwa,Caron-Huot:2014lda} that in many of these cases 
a  basis of MIs
can be found, such that the system of differential equations takes a particularly
simple form, commonly referred to as \textsl{canonical form}. The system is said to be in canonical
form if the regularisation parameter $d$ can be completely factorised from the kinematics, appearing
as an explicit $(d-4)$ factor in front of the matrix of the system. In addition, the coefficients of the matrix
must be total differentials of logarithms of functions of the external invariants 
(i.e. they are said to be in d-log form). 
A canonical basis is particularly convenient as it allows a straightforward integration 
as series expansion in $(d-4)$
and, due to the d-log form of the coefficients, it integrates directly to MPLs of uniform transcendental 
weight. Criteria for the construction of candidate canonical integrals
have been presented in~\cite{Henn:2013pwa} and
developed in detail, for example, in~\cite{Bern:2014kca}. 
In the special cases in which
the differential equations depend only linearly on the dimensions $d$, the Magnus
algorithm can be used to perform a rotation of the system to a canonical form~\cite{Argeri:2014qva}.
For a recent application of the algorithm see~\cite{DiVita:2014pza}.
A completely different approach based on Moser algorithm~\cite{Moser} has been developed
for the univariate case in~\cite{Lee:2014ioa} and discussed also independently in~\cite{Henn:2014qga}. 
Another interesting approach is based on the properties of higher order
differential equations fulfilled by the individual master integrals~\cite{Hoschele:2014qsa}.
In spite of all this impressive progress, a fully
automated algorithm, working also in the multivariate case, is still missing.
It has nevertheless become clear that, if one can find a basis of MIs 
whose differential equations become triangular as $d \to 4$, it is often (\textsl{but not always})
relatively easy to bring it in canonical form by removing the undesired terms in 
the differential equations~\cite{Gehrmann:2014bfa}. 

Indeed, different examples are known where neither finding a canonical basis nor 
a triangular one is possible. In all these cases
the master integrals cannot be integrated in terms of simple 
multiple polylogarithms only~\cite{Laporta:2004rb,Adams:2013nia,Remiddi:2013joa,Adams:2014vja}.
It becomes therefore a very interesting problem that of finding a set of criteria to determine 
whether, given a Feynman graph, there exists a basis of MIs
whose system of differential equations becomes triangular in the limit for $d \to 4$.
This would provide a way to easily classify all possible diagrams where this is not possible and that
would therefore be expected to evaluate to more complicated classes of function.
Aim of this paper is to show that a large amount of information about the possibility of 
achieving such decoupling can be extracted by studying the IBPs for fixed
even numbers of dimensions, i.e. in the limit $d \to 2\,n$,
where $n$ is any integer number. 
Tarasov-Lee shifting identities~\cite{Tarasov:1996br,Lee:2009dh} 
allow, in fact, to directly relate the structure of the differential equations in any even number of
dimensions with the physical case $d=4$.
Indeed, since Feynman integrals are usually divergent at $d=2\,n$, this limiting procedure must
be carried out as a Laurent series in $(d-2\,n)$. 

The rest of the paper is organised as follows. 
In Section~\ref{sec:IBPs} we review the use of integration by parts identities for the reduction to master
integrals and we summarise the main results of the differential equations method. 
In Section~\ref{sec:IBPsN}
we outline the central idea of the paper. 
We show in particular what kind of information can be extracted by solving the IBPs as Laurent series 
for $d \to 2\,n$ and how to use it to simplify the system of differential equations.
In Section~\ref{sec:Ex} we then apply these ideas explicitly to many different examples of increasing complexity.
Some comments on the method are given in Section~\ref{sec:Comments}, where we try as well to point out
some relevant open issues.
Finally we conclude in Section~\ref{sec:Concl}.
In Appendix~\ref{App:Sch} we compare our method with the Schouten identities 
introduced in~\cite{Remiddi:2013joa},
while in Appendix~\ref{App:Dim} we show how to shift a system of differential equations of an even number of
dimensions.

\section {Integration by parts identities and differential equations} \labbel{sec:IBPs} \setcounter{equation}{0} 
\numberwithin{equation}{section}

Let us consider an $l$-loop scalar Feynman integral depending on $P$ independent
external momenta $p_i$
\begin{align}
\I(d;a_1,...,a_\tau,b_1,...,b_\sigma) = 
\int \prod_{j=1}^l\, \frac{d^d k_j}{(2 \pi)^2}\,
\frac{S_1^{b_1}...\, S_\sigma^{b_\sigma}}{D_1^{a_1}...\, D_\tau^{a_\tau}} \,, \labbel{eq:scalint}
\end{align}
where $k_i$ are the loop momenta, $D_i=q_i^2-m_i^2$ are the propagators 
and $S_i = k_j \cdot p_l$ are irreducible scalar products. In view of the discussion below, we 
will usually denote the integrals of a topology as in~\eqref{eq:scalint}, keeping explicitly only the dependence on the dimensions $d$ and on the powers of denominators
and scalar products.
In dimensional regularisation, for every integral of the form~\eqref{eq:scalint} there 
always exists a value of the space-time dimensions $d$, such that
the integral is convergent\footnote{All scaleless integrals in dimensional regularisation are
zero for consistency.}. Necessary condition for the convergence of an integral is the integrand 
be zero at the boundaries. This condition can be mathematically rephrased as
\begin{align}
\int \prod_{j=1}^l\, \frac{d^d k_j}{(2 \pi)^2}\, \frac{\partial}{\partial k_n^\mu}\left( v_m^\mu\, 
\frac{S_1^{b_1}...\, S_\sigma^{b_\sigma}}{D_1^{a_1}...\, D_\tau^{a_\tau}} \right) = 0 \,, \label{eq:ibps}
\end{align}
where the $v_m^\mu$ are any of the external or internal momenta 
 $v_m^\mu = \left\{ k_1,...,k_l; p_1,...,p_P \right\}.$ Such an identity is called
 an integration by part identity or IBP.
It is clear that in this way
$l(l+P)$ IBPs can be established for each integrand.
Upon explicitly evaluating the derivatives and contracting with the
momenta $v_m^\mu$, new integrals belonging to
the same topology (i.e. integrals with the same set of denominators) are generated.
In particular, each IBP identity can relate integrals with $(s-1)$,
$s$ and $(s+1)$ powers of scalar products, and
$(t+r)$ or $(t+r+1)$ powers of propagators.
Notice that, by contracting with $v_m^\mu$, new reducible scalar products
can be generated, which could then simplify some of the denominators
producing integrals belonging to any of the $(t-1)$ sub-topologies of the 
original graph.

It has been shown~\cite{Laporta:1996mq,Gehrmann:1999as,Laporta:2001dd} that
the system of IBPs, which appears in general to be over-constrained, can instead be solved
allowing to express most of the integrals as linear combinations of a small subset of basic integrals,
dubbed \textsl{master integrals} (MIs). Indeed, as for any
algebraic basis, the choice
is not unique and, by suitably changing the basis, one can substantially simplify
the calculation of the integrals, as we will discuss later in this paper.\footnote{A different
approach to reduction to MIs using hyperelliptic curves, and its equivalence to the IBPs
in some explicit cases, was recently discussed in~\cite{Georgoudis:2015hca}.}

Let us consider now a Feynman graph (or a topology) characterised by a set of propagators $D_i$
and irreducible scalar products $S_i$. All integrals will depend on the space-time
dimensions $d$ and on the external invariants $x_{ij} = p_i \cdot p_j$, where $p_i$ are
as usual the external momenta.
Let us assume, for definiteness, that by solving
the system of IBPs\footnote{Note that, in order to have a complete reduction, one must
also consider all possible symmetry relations among the integrals due to shifts of the loop momenta.}
all integrals for the given
graph can be expressed in terms of a basis of $N$ independent MIs $\I_i(d;x_{ij})$
with $i=1,...,N$, which are of the same form of Eq.~\eqref{eq:scalint}.
Differentiating with respect to any of the external invariants $x_{ij}$ amounts to differentiating
with respect to linear combinations of the external momenta $p_i^\mu$~\cite{Remiddi:1997ny}.
Therefore, by acting with these differential operators 
directly on the \textsl{integrands} of~\eqref{eq:scalint}, one produces
linear combinations of integrals belonging to the same Feynman graph and to
its sub-topologies. The latter can again be reduced to MIs, 
generating in this way a system of
$N$ linear first order differential equations with rational coefficients in any of the invariants
$x_{ij}$. Suppressing the dependence on the sub-topologies, which can be considered as a  
known inhomogeneous term in a bottom-up approach, the homogeneous part of the 
system can always be written as

\begin{align}
\frac{\partial}{\partial x_{ij}} \left( \begin{array}{c} \I_1(d;x_{ij}) \\ ... \\ \I_N(d;x_{ij}) \end{array}\right)
&= \left( \begin{array}{ccc} c_{11}(d;x_{ij}) & ... & c_{1N}(d;x_{ij})\\
... & ... & ... \\
 c_{N1}(d;x_{ij}) & ... & c_{NN}(d;x_{ij})
 \end{array}\right) 
\left( \begin{array}{c} \I_1(d;x_{ij}) \\ ... \\ \I_N(d;x_{ij}) \end{array} \right)\,,
\label{eq:sysdeq}
\end{align}
where the coefficients $c_{ij}(d;p^2)$ are simple rational functions of the dimensions $d$ and of the 
external invariants $x_{ij}$. We can rewrite the system in matrix form as
\begin{align}
\frac{\partial}{\partial x_{ij}} \, \vec{\I}(d;x_{ij}) = A(d;x_{ij})\, \vec{\I}(d;x_{ij})\,, \labbel{eq:diffeqgen}
\end{align}
where we introduced the vector of master integrals $\vec{\I}(d;x_{ij})$ and the matrix of
the coefficients $A(d;x_{ij})$.

The system~\eqref{eq:sysdeq} is, in the general case, coupled and can
therefore be rephrased as an $N$-th order differential equation for any of the MIs 
$\I_i(d;x_{ij})$.
In most practical applications we are  interested in determining the  MIs
as Laurent expansion for $d \approx 4$
\begin{align}
&\I_{i}(d;x_{ij}) = \sum_{\alpha=-a}^\infty \, \I_{i}^{(\alpha)}(4;x_{ij}) \, (d-4)^{\alpha}\,. \labbel{eq:ser4}
\end{align}
By expanding both left- and right-hand side of~\eqref{eq:sysdeq} one is left with a chained
system of $N$ differential equations where, at any order $\alpha$, the previous orders can only appear
as inhomogeneous terms.

\subsection{An optimal basis of master integrals}

It has been shown by Tarasov and Lee~\cite{Tarasov:1996br,Lee:2009dh} that the value of a
Feynman integral in $d$ space-time dimensions can be directly 
related to that of the same
integral in $d-2$ or $d+2$ space-time dimensions. This implies that, if all MIs of a given graph are known
as Laurent expansion in any \textsl{even} number of dimensions, $d =2\,n$,
\begin{align}
&\I_{i}(d;x_{ij}) = \sum_{\alpha=-b}^\infty \, \I_{i}^{(\alpha)}(2\,n;x_{ij}) \, (d-2\,n)^{\alpha}\,, \labbel{eq:ser2n}
\end{align} 
then the
coefficients of their series expansions in $d = 4$, $\I_{i}^{(\alpha)}(4;x_{ij})$ in~\eqref{eq:ser4}, can be obtained
as linear combinations of the $\I_{i}^{(\alpha)}(2\,n;x_{ij})$. For more details
see for example~\cite{Laporta:2004rb,Remiddi:2013joa}
and the discussion in Appendix~\ref{App:Dim}.
 
Indeed, changing the basis of MIs changes the form of the matrix $A(d;x_{ij})$
in equation~\eqref{eq:diffeqgen}. 
An interesting problem is therefore how to define an \textsl{optimal} basis of MIs in order to
simplify as much as possible the system~\eqref{eq:sysdeq}.
Since
we are interested in computing the MIs as Laurent expansion in $(d-4)$ (or, in general, in $(d-2\,n)$), an
obvious simplification would occur if we could decouple some of the differential equations,
at least in the considered limit. In particular, given a system of $N$ coupled equations, one could think of 
classifying the complexity of the latter by determining the \textsl{minimum number} of differential equations that cannot 
be decoupled in the limit $d \to 4$ (or more generally  $d \to 2\,n$). 

At this point it is useful clarify more precisely what we mean with
\textsl{decoupling} in this context.
Let us consider a $2 \times 2$ coupled system of
differential equations\footnote{Again, we neglect the inhomogeneous terms everywhere.}
\begin{align}
\frac{\partial}{\partial x} \, \vec{\I}(d;x) = A(d;x)\, \vec{\I}(d;x)\,,
\end{align}
where  $\vec{\I}(d;x) =  \left( I_1(d;x), I_2(d;x) \right)$ is the 2-vector of  unknown functions,
$A(d,x)$ is a $2 \times 2$ matrix, $d$ are the space-time
dimensions and $x$ is a variable the functions depend on\footnote{In the case of Feynman integrals
$x$ represents a generic Mandelstam variable.}.
Assume now for simplicity that the functions $I_1(d;x), I_2(d;x)$ are finite in the limit $d \to 4$
and that the matrix $A(d;x)$ does not contain any explicit poles in $1/(d-4)$.
Assume finally that, in the limit $d \to 4$, the matrix $A(d;x)$ has \textsl{non-zero} non-diagonal
entries, and therefore that the system is coupled in the limit $d \to 4$. Of course, by expanding 
the entries of $A(d;x)$ in $(d-4)$ we can write our system as

\begin{align}
\frac{\partial}{\partial x} \, \vec{\I}(d;x) = A^{(0)}(4;x)\, \vec{\I}(d;x) + (d-4) A^{(1)}(4;x)\, \vec{\I}(d;x) 
+ \mathcal{O}\left( (d-4)^2 \right)\,.
\end{align}
It is now clear that if, by any means, we can find two independent solutions to the $2 \times 2$ system
\begin{align}
\frac{\partial}{\partial x} \, \vec{f}(x) = A^{(0)}(4;x)\, \vec{f}(x)\,,
\end{align}
say $ \left( v_1(x), v_2(x)\right)$ and $ \left( w_1(x), w_2(x)\right)$,
then we can define the new vector $\vec{\J}(d;x)$ through the rotation
\begin{equation}
\vec{\J}(d;x) = G(x)\,\vec{\I}(d;x)\,, \quad \mbox{with} \quad 
G(x) = \left( \begin{array}{cc} v_1(x) & w_1(x) \\ v_2(x) & w_2(x)  \end{array}\right)\,,
\end{equation}
such that the differential equations satisfied by $\vec{\J}(d;x)$ assume the form
\begin{align}
\frac{\partial}{\partial x} \, \vec{\J}(d;x) = (d-4) \, G^{-1}(x)\,A^{(1)}(4;x) G(x) \, \vec{\I}(d;x) 
+ \mathcal{O}\left( (d-4)^2 \right)\,,
\end{align}
i.e. they become trivial in the limit $d \to 4$.
The matrix $G(x)$ can be of course \textsl{arbitrarily complicated}, as it contains the solutions of a second
order differential equation. In this case we would have of course achieved a decoupling, but at the price of
having to solve a coupled system of differential equations, which is in the general case not possible.

On the other hand, what we are really interested in is to determine whether a basis of MIs exists such that
some of the non-diagonal terms of the matrix $A(d;x)$ become zero in the limit $d \to 4$, and such that this
basis can still be reached from our starting basis only through IBPs (i.e. without having to solve
a coupled system of differential equations!). What this means in practice is 
that, if such a basis existed, then the rotation matrix $G$ would assume a very simple form, 
namely it would contain only
rational functions of the external invariants $x_{ij}$ (and of the dimensions $d$). 
This new basis would therefore fulfil a system of differential
equations where some (or all) of the MIs decouple in the limit $d \to 4$,
and still it would be a system of linear differential equations with rational coefficients only.
In this respect we note that, for all known cases of MIs which can be integrated
in terms of multiple polylogarithms, a change of basis in the sense described above can be found
and the system of differential equations can be put in
\textsl{triangular} form as $d \to 4$
\begin{align}
\frac{\partial}{\partial x_{ij}} \vec{\J}(d;x_{ij})
= T(4;x_{ij}) \vec{\J}(d;x_{ij})+ \mathcal{O}(d-4)\,,
\end{align}
where $T(4;x_{ij})$ is a triangular matrix and does not depend on the dimensions $d$. 
From the point of view of the classification outlined above this corresponds to the easiest
case, where all equations decouple in the limit $d \to 4$ and, effectively, the problem
reduces to a series of independent integrations by quadrature.
Finding a basis in this form is often a first step towards
a canonical basis in the sense introduced in~\cite{Henn:2013pwa}.
We want to stress here that of course all these considerations apply in the very same
way for any integer number of dimensions $d \to n$ (even or odd).

Unfortunately a change of basis of this kind cannot always be found.
Several cases are known where the system cannot be \textsl{completely triangularised}
and instead at least two differential equations remain coupled.
In these cases MPLs turn out not to be enough for describing the solution and the
class of functions must be enlarged to include also elliptic generalisations of the latter~\cite{Broedel:2015hia}.
It is unclear whether this will be the end of the story, since cases where three or more coupled
equations survive are relatively easy to find, as we will show in the following.
What still appears to be missing is a criterion to determine, given a Feynman graph, what is the
minimum number of equations which cannot be decoupled. 
Together with simplifying as much as possible the problem at hand, 
this could also give a hint to which class
of functions are required for describing the solution.

\section{Reading the IBPs in fixed numbers of dimensions} \labbel{sec:IBPsN} \setcounter{equation}{0} 
\numberwithin{equation}{section}
In order to find a possible working criterion to determine the minimum number of coupled differential 
equations we should go back to think how the differential equations are derived.
We saw that differentiating a master integral with respect to the external invariants produces
new integrals belonging to the same Feynman graph. By using the IBPs one can then
reduce these integrals to MIs, ending up with a system of differential equations. 
If we start with $N$ master integrals we will obtain in general 
a coupled system of $N$ linear differential equations. 
The fact that the $N$ differential equations
are coupled can be seen, in this respect, as due to the linear independence of the
$N$ master integrals in $d$ dimensions. 

As we already discussed, for any physical application we are interested in computing Feynman
integrals as Laurent series in $(d-4)$ or, more generally, in $(d-2\,n)$ with $n \in \N$.
Of course, different integrals have different
degrees of divergence, i.e. their Laurent expansion starts at different orders in $(d-2\,n)$.
For any value of the dimensions, nevertheless, the maximal divergence can be computed 
in dimensional regularisation and
depends only on the topology of the graph under consideration (i.e. on the number of 
loops, of external legs etc.).
We can therefore imagine to first generate the IBPs in $d$ dimensions and 
then expand them as a Laurent series in $(d-2\,n)$, obtaining in this way a chained set of 
systems of IBPs, one for every order in $(d-2\,n)$.
It is clear that, by construction, at every order in $(d-2\,n)$, the homogeneous part of each system 
will be identical, while the inhomogeneous part will contain the previous orders of the expansion
(and the sub-topologies, that we will neglect throughout).
If we limit ourselves to the first order of the expansion,
i.e. the one corresponding to the highest pole in $(d-2\,n)$, the system of equations that we are left with is equivalent
to the original system of IBPs where $d$ is fixed to be $d=2\,n$, and corresponds to the
sole homogeneous system.

Now, it is very well known that upon fixing the number of space-time dimensions in the IBPs to an integer value
it may happen that some of the equations degenerate and, in particular, that some of the integrals that
used to be linearly independent for generic values of $d$, become linearly dependent from each other.
From the point of view of the differential equations satisfied by the master integrals, if some of the integrals
were to become linearly dependent in the limit $d \to 2\,n$, one would expect that those masters should
not bring any new information in that limit and it should therefore be possible to
\textsl{decouple} them from the system of differential equations as $d \to 2 \, n$. 
Let us try to state this point more precisely. As exemplification we consider a topology that is
reduced to 2 master integrals which we call $\I_1(d;x)$ and $\I_2(d;x)$, where $d$ are the dimensions
and $x$ is a generic Mandelstam variable. Neglecting the sub-topologies 
the system of differential equations that they satisfy can be written as
\begin{equation}
 \left\{ \begin{array}{c} \frac{\partial}{\partial \, x}\,\I_1(d;x) = c_{11}(d;x)\,\I_1(d;x) + c_{12}(d;x) \, \I_2(d;x) \\ \\
                                    \frac{\partial}{\partial \, x} \I_2(d;x) =  c_{21}(d;x)\,\I_1(d;x) + c_{22}(d;x) \, \I_2(d;x)
\end{array} \right.\,,
 \labbel{eq:sysgen}
\end{equation}
where the $c_{ij}(d;x)$ are rational functions.
Let us now follow the argument above and generate the IBPs fixing $d=2\,n$. Let us assume that, by solving
this simplified system, one of the two master integrals becomes linearly dependent from the other one and
the new IBPs produce the relation
\begin{equation}
\I_2(2\,n;x) = b(x)\,\I_1(2\,n;x)\,, \labbel{eq:relgen}
\end{equation} 
where $b(x)$ is a rational function of the Mandelstam variables\footnote{To be precise we should recall that, since the master integrals can be divergent, this relation cannot be seen, in general,
 as a real relation between the two masters.}.
Equation~\eqref{eq:relgen} implies that in $d=2\,n$ one of the two master integrals becomes linearly
dependent in the sense of the IBPs. 
According to the argument above we would therefore expect to be able to
decouple the two differential equations in this limit. In order to see this it is useful to ask ourselves how
such a relation can emerge from the original $d$-dimensional IBPs. Let us imagine that upon
solving the IBPs for generic $d$, we can find a $d$-dimensional relation expressing a given integral 
of the graph under consideration,
say $K(d;x)$, as a linear combination of the two masters and such that
\begin{equation}
K(d,x) = \frac{1}{d-2\,n} \left( b_1(d;x)\,\I_1(d;x) + b_2(d;x)\,\I_2(d;x)   \right)\,, \labbel{eq:relgenibps}
\end{equation}
with $b_1(x)/b_2(x)= b(x)$ and $\lim_{d \to 2\,n} b_i(d;x) = b_i(x)$, for $i=1,2$. 
It is clear that, if this is the case, the IBPs which would generate this identity for 
generic $d$, would instead generate~\eqref{eq:relgen} once $d$ is fixed to be $d=2\,n$.
These relations are precisely what we are looking for.
To refer to the latter we will often use throughout the paper the notation
\begin{equation}
b_1(d;x)\,\I_1(d;x) + b_2(d;x)\,\I_2(d;x)  = \mathcal{O}(d-2\,n),
\end{equation}
or equivalently
\begin{equation}
b_1(x)\,\I_1(d;x) + b_2(x)\,\I_2(d;x)  = \mathcal{O}(d-2\,n),
\end{equation}
where it should be understood that, in general, this does not mean that the combination above is really of
order $\mathcal{O}(d-2\,n)$, but simply that it \textsl{becomes zero upon setting $d=2\,n$ in the IBPs.}
Note that, of course, using the $b_i(x)$ instead of the $b_i(d;x)$ can only produce corrections of order 
$\mathcal{O}(d-2\,n)$ due to~\eqref{eq:relgenibps}. 
We will see many examples of these relations in the sections below.

Naively, the fact that only one integral is linearly independent for $d=2\,n$ would require that the integral
itself should satisfy a first order differential equation as $d \to 2\,n$. Finding a basis of master integrals
for which this first order equation emerges is equivalent to finding a basis which decouples 
the system~\eqref{eq:sysgen}.
To this aim let us perform the following rotation of the master integral basis 
\begin{equation}
\J_1(d;x) =  b_1(x)\,\I_1(d;x) + b_2(x)\,\I_2(d;x)  \,,\qquad \J_2(d;x) = \I_2(d;x)\,.\labbel{eq:rotation}
\end{equation} 
The system~\eqref{eq:sysgen} under this rotation becomes
\begin{align}
&\frac{\partial}{\partial \, x}\,\J_1(d;x) = 
\left( c_{11}(d;x) + \frac{b_2(x) c_{21}(d;x) + b_1'(x)}{b_1(x)} \right) \,\J_1(d;x) \nonumber \\
&\quad + \left( b_1(x)c_{12}(d;x) + b_2(x) \left( c_{22}(d;x)-c_{11}(d;x) \right) + b_2'(x) 
- \frac{b_2(x) \left[ b_2(x) c_{21}(d;x) + b_1'(x) \right]}{b_1(x)}\right) \,\J_2(d;x)
\nonumber \\ & \nonumber \\
&\frac{\partial}{\partial \, x}\,\J_2(d;x) = \left( c_{22}(d;x) - \frac{b_2(x)}{b_1(x)} c_{21}(d;x) \right) \J_2(d;x) 
+ \frac{c_{21}(d;x)}{b_1(x)} \J_1(d;x)\,. \labbel{eq:gendecoupling}
\end{align}
Equations~\eqref{eq:gendecoupling} do not look particularly illuminating at first glance.
We claim nevertheless that these equations are precisely what we were looking for. 
The basis $\J_1(d;x), \J_2(d;x)$ defined in~\eqref{eq:rotation}, in fact, has been chosen 
in order to exploit the linear dependence of the two master integrals
in the limit $d \to 2\,n$. In this limit the IBPs tell us that $\J_1(d;x)$ is by construction
suppressed by a factor $(d-2\,n)$ and therefore decouples from the problem. 
We expect therefore that the differential equation for the latter should decouple
in this limit or, in other words, that
\begin{equation}
\left( b_1(x)c_{12}(d;x) + b_2(x) \left( c_{22}(d;x)-c_{11}(d;x) \right) + b_2'(x) 
- \frac{b_2(x) \left[ b_2(x) c_{21}(d;x) + b_1'(x) \right]}{b_1(x)}\right) \propto \mathcal{O}(d-2\,n)\,.
\labbel{eq:gensuppr}
\end{equation}
If this is true then upon expanding the system of differential equations as Laurent series
in $(d-2\,n)$ one can, at every oder, first solve the differential equation for $\J_1(d;x)$
by quadrature, and then use this as an input for the second equation.
A rigorous mathematical proof of equation~\eqref{eq:gensuppr} is outside the scope of this paper
and we will limit ourselves to show explicitly how this works in practice with several
examples of different complexity.

The considerations above can be easily generalised to $N$ master integrals
$\I_1$,..., $\I_N$.
In this case one starts with a system of $N$ coupled differential equations. By solving the IBPs for $d=2\,n$
one can then verify how many of the master integrals become linearly dependent in this limit.
Assuming that $N-M$ integrals remain independent, this means that $M$ relations like~\eqref{eq:relgenibps}
can be found, say
\begin{alignat}{3}
&K_1(d;x) &=& \frac{1}{d-2\,n}\left( b_{11}(d;x) \I_1(d;x) + ... + b_{1N}(d;x)\I_N(d;x) \right) \nonumber \\
&...&& \nonumber \\
&K_{M}(d;x) &=& \frac{1}{d-2\,n}\left( b_{M1}(d;x) \I_1(d;x) + ... + b_{MN}(d;x)\I_N(d;x) \right)\,, \labbel{eq:relgenibpsN}
\end{alignat}
and the $b_{ij}(d;x)$ are as always rational functions of the dimensions and of the Mandelstam 
variables\footnote{For this to be true the relations~\eqref{eq:relgenibpsN} must be linearly independent
in the limit $d \to 2\,n$.}.
As for the previous example we will often write these relations as
\begin{align}
&b_{11}(d;x) \I_1(d;x) + ... + b_{1N}(d;x)\I_N(d;x) = \mathcal{O}(d-2\,n) \nonumber \\
&... \nonumber \\
&b_{M1}(d;x) \I_1(d;x) + ... + b_{MN}(d;x)\I_N(d;x) = \mathcal{O}(d-2\,n)\,, \labbel{eq:genrelN}
\end{align}
where once more we imply that these combinations become zero upon setting $d=2\,n$ in the IBPs.
As before we define $b_{ij}(x) = \lim_{d \to 2\,n} b_{ij}(d;x)$ and, following the same reasoning, 
we can then try to rotate the basis of master integrals to

\begin{alignat}{3}
&\J_1(d;x) &&=\; && b_{11}(x) \I_1(d;x) + ... + b_{1N}(x)\I_N(d;x) \nonumber \\
& .. \nonumber \\
&\J_M(d;x) &&=\; && b_{M1}(x) \I_1(d;x) + ... + b_{MN}(x)\I_N(d;x) \nonumber \\
&\J_{M+1}(d;x) &&=\; && \I_{M+1}(d;x) \nonumber \\
& .. \nonumber \\ 
&\J_{N}(d;x) &&=\; && \I_{N}(d;x)\,. \labbel{eq:rotationN}
\end{alignat}
Under the rotation~\eqref{eq:rotationN}, we expect the $M$ integrals $\J_1(d;x)$, ..., $\J_M(d;x)$
to decouple from the remaining independent integrals in the limit $d \to 2\,n$, as in~\eqref{eq:gendecoupling}.
One must be cautious here on what is intended by decoupling. 
According to the arguments above, upon the change of basis~\eqref{eq:rotationN}, 
we expect the system of differential equations to split into two blocks in
the limit $d \to 2\,n$, one $M \times M$ and the other $(N-M) \times (N-M)$.
This would correspond, order by order in $(d-2\,n)$, to an $M$-th
plus an $(N-M)$-th order differential equation, unless for some other reason internally the
two blocks of differential equations further decouple in this limit.
On the other hand, for the Feynman graphs that we considered so far
(see for example Sections~\ref{sec:sun2} and~\ref{sec:ban}), even a stronger claim can be made.
In these cases the rotation~\eqref{eq:rotationN} not only splits the system into two blocks,
as described above, but it also produces an explicit $(d-2\,n)$ in front of the whole $M \times M$
block originating from relations~\eqref{eq:genrelN}. This explicit overall factor allows to effectively 
reduce the problem to the solution of one single $(N-M)$-th differential equation, plus 
$M$ integrations by quadrature. 
The reason for this behaviour is still partly unclear and deserves further study.

Summarising, the discussion above brings us to the following conclusion. Given a topology with $N$
master integrals which fulfil a set of $N$ coupled differential equations in $d$ space-time dimensions,
the study of the IBPs in fixed numbers of dimensions, say $d=2\,n$, provides a tool to determine
how many master integrals can be decoupled from the differential equations as $d \to 2\,n$.
Of course, the arguments given above are partly oversimplified and we have not provided here any rigorous 
mathematical proof. The structure of the differential equations can be in general very involved
and, instead of embarking on complicated mathematical arguments, we prefer to show explicitly how this
ideas can be simply applied to different cases of increasing complexity. In the next section we will start off 
by considering simple examples where, by fixing the number of dimensions to an even integer value, only
one master integral remains linearly independent and therefore the problem can be reduced to the solution
of one linear differential equation. We will then move to more interesting cases where, 
even in fixed numbers of dimensions, more than one master integral remain linearly independent and 
one cannot avoid the problem of solving higher order differential equations which give rise to more
complicated mathematical structures.

\section{Explicit examples} \labbel{sec:Ex} \setcounter{equation}{0} 
\numberwithin{equation}{section} 
In the previous section we outlined the main ideas behind this paper.
We argued that the IBPs might degenerate
in the limit of fixed (even) integer numbers of dimensions $d \to 2 \, n$, 
such that some of the master integrals
become effectively linearly dependent from each other. While this is a very well known fact,
we argued that this degeneracy, if present, can be used in order to simplify
the system of differential equations satisfied by the master integrals. 
In this section we will present many explicit examples of this simple idea. 
We will start by studying the two-loop
sunrise graph with one massive and two massless propagators, Section~\ref{sec:sun1}, and a two-loop
triangle with three off-shell legs, Section~\ref{sec:triangle}. In both examples there are only two master integrals
and by studying the IBPs in fixed even numbers of dimensions, one relation can be found, allowing to 
decouple the differential equations in that limit. 
We will then consider the case of the two-loop massive sunrise, Section~\ref{sec:sun2}, and of a non-planar
two-loop triangle, Section~\ref{sec:crossed}. In both cases not all equations can be decoupled, and a 
minimal bulk of two differential equations remains coupled, giving rise to elliptic functions. 
We will then study the case
of a two-loop massive triangle with three master integrals, Section~\ref{sec:triangle2}. 
Here, similarly to the non-planar two-loop triangle, there are 
three master integrals. In this case, nevertheless, the differential equations can be completely decoupled
and the solution can be written in terms of MPLs.
Finally, as a last example,
we will move to the three-loop massive banana graph~\ref{sec:ban}. In this case, we will study different possible
mass-arrangements of increasing complexity, showing how the number of master integrals changes consequently,
and how our method allows to determine easily which subset of master integrals can be immediately
decoupled from the differential equations.

\subsection{The two-loop sunrise with one massive propagator} \labbel{sec:sun1} \setcounter{equation}{0} 
\numberwithin{equation}{section}
Let us start off by considering the case of the
two-loop sunrise with one massive and two massless propagators. We define the following 
set of integrals belonging to its Feynman graph
\begin{align}
I(d;n_1,n_2,n_3,n_4,n_5) &= \Sunriseone{p^2} \nonumber \\ &=
\int \D^d k  \D^d l\, \frac{(k \cdot p)^{n_4}(l \cdot p)^{n_5}}
{\left( k^2 \right)^{n_1} \left( l^2 \right )^{n_2} \left((k-l+p)^2-m^2 \right)^{n_3}}\,, \labbel{eq:sunr1}
\end{align}
where $p^2 = s$ is the momentum transfer. The integration measure is defined as 
\begin{equation}
\D^d k = C(d)\, \frac{d^dk}{(2 \pi)^d}\,, \labbel{eq:measure}
\end{equation}
and the explicit form of the function $C(d)$ is not relevant for the considerations below.
Note that this Feynman graph does not
contain any sub-topology.
We keep explicit only the dependence on the space-time dimensions $d$ and on the 
powers of the denominators and scalar products, which will be important for what follows.
Performing a usual reduction through IBPs one finds two independent MIs, which can
be chosen as
\begin{equation}
\I_1(d;s) = I(d;1,1,1,0,0)\,, \qquad  \I_2(d;s) = I(d;1,1,2,0,0)\,. \labbel{eq:missun1}
\end{equation}

Using the methods outlined in the previous sections we can now derive the differential equations
fulfilled by $\I_1$ and $\I_2$ in the momentum squared $s$. This step can be performed automatically
using, for example, Reduze 2~\cite{vonManteuffel:2012np}, and we end up with the following
$2 \times 2$ linear system
\begin{align}
&\frac{d \I_1}{d s} = \frac{(d-3)}{s} \, \I_1 - \frac{m^2}{s}\, \I_2 \nonumber \\
&\frac{d \I_2}{d s} = \frac{(d-3)(3 d - 8)}{2\,m^2} \,
\left(\frac{1}{s} - \frac{1}{s-m^2}\right) \I_1 
+ \left(\frac{2(d-3)}{s-m^2} - \frac{(3d-8)}{2\,s} \right)\, \I_2\,. \labbel{eq:deqsun1}
\end{align}
As one can immediately see, the equations are coupled for any \textsl{even} value of the
dimensions $d$\footnote{On the other hand, the equations become triangular as $d \to 3$.}.

It is well known that these integrals can be computed as a series expansion in $d \to 4$
in terms of HPLs only, see for example~\cite{Huber:2015bva}. 
Let us then try and use the ideas outlined in Section~\ref{sec:IBPsN} in order
to decouple the differential equations in the limit $d \to 4$. 
First of all note that, in the limit $ d \to 4$, both master 
integrals are UV divergent and in particular they both develop a double pole
\begin{align}
&\I_1(d;s) = \frac{1}{(d-4)^2}\,\I_1^{(-2)}(4;s) + \frac{1}{(d-4)}\,\I_1^{(-1)}(4;s) + \mathcal{O}(1) \\
&\I_2(d;s) = \frac{1}{(d-4)^2}\,\I_2^{(-2)}(4;s) + \frac{1}{(d-4)}\,\I_2^{(-1)}(4;s) + \mathcal{O}(1)\,.
\end{align}

Moreover it is easy to show that any integral of the form~\eqref{eq:sunr1} can \textsl{at most} 
develop a double pole in $(d-4)$.
Equipped with these consideration, let us now produce the
IBPs for this Feynman graph as described above but, instead of solving them
keeping the full dependence on the parameter $d$,
we can set $d=4$.\footnote{The possibility
of solving IBPs for fixed values of $d$ is already implemented in the development version of
Reduze 2.} As we discussed in detail in the previous section, this is equivalent to expanding the IBPs
in Laurent series, and considering the first of the chained systems of equations obtained, 
namely the one corresponding to the double pole in $(d-4)$.
Following the arguments of the previous section, we would expect to find a degeneracy of the
two master integrals in $d=4$, which should then allow us to decouple the two differential equations.
As expected the two masters~\eqref{eq:missun1} become linearly dependent
\begin{equation}
\I_2^{(-2)}(4;s) = \frac{1}{m^2} \I_1^{(-2)}(4;s) \,. \labbel{eq:relsun1d4}
\end{equation}
As discussed in Section~\ref{sec:IBPsN}, such a relation must come from a corresponding
$d$-dimensional IBP. Indeed, if one considers the original $d$-dimensional system of IBPs
and solves it for the two masters, it is easy to find the following relation
\begin{align}
I(d; 2,1,1,0,0) = 
- \left(\frac{1}{d-4}\right)\, \frac{(d-3)}{s-m^2}  \left(\,(3d-8)\I_1(d;s)-4\,m^2\,\I_2(d;s) \right)\,.
\labbel{eq:ibpd4}
\end{align}
In the limit $d \to 4$ Eq.~\eqref{eq:ibpd4} trivially generates Eq.~\eqref{eq:relsun1d4}.
In the notation of Section~\ref{sec:IBPsN} we can write this relation as
\begin{equation}
(3d-8)\I_1(d;s)-4\,m^2\,\I_2(d;s) = \mathcal{O}(d-4)\,,
\end{equation}
or, equivalently, keeping also in the right-hand side only terms of $\mathcal{O}(d-4)$,
\begin{equation}
 \I_1(d;s) - m^2\, \I_2(d;s) = \mathcal{O}(d-4)\,,
\end{equation}
recalling that this does not mean that this linear combination is of order $\mathcal{O}(d-4)$,
but that it becomes zero if we fix $d=4$ in the IBPs.

In this particular case, since the Feynman graph under consideration does not have any sub-topologies,
Eq.~\eqref{eq:relsun1d4} can be seen as a \textsl{real relation}
between the highest poles of the two master integrals. 
This relation, which is naturally derived from the IBPs only,
can be easily verified by computing the highest poles of the two master integrals.
A very simple exercise gives 
\begin{align}
& \I_1(d;s) = \frac{1}{(d-4)^2} \left( \frac{m^2}{2}\right) + \mathcal{O}\left( \frac{1}{(d-4)} \right)\nonumber \\
& \I_2(d;s) = \frac{1}{(d-4)^2} \left( \frac{1}{2}\right) + \mathcal{O}\left( \frac{1}{(d-4)} \right)\,, \labbel{eq:ressun4}
\end{align}
in agreement with Eq.~\eqref{eq:relsun1d4}. The overall normalisation of Eq.~\eqref{eq:ressun4}
is of course arbitrary and it has to do with the choice for the integration measure~\eqref{eq:measure}.
Let us now try and exploit this relation in order to simplify the system of differential
equations~\eqref{eq:deqsun1}.
We perform the change of basis from the ``standard'' MIs $\I_1(d;s)$, $\I_2(d;s)$, to the new
MIs defined as
\begin{align}
&\J_1(d;s) = \I_1(d;s) - m^2\, \I_2(d;s)\,,  \qquad
\J_2(d;s) = \I_1(d;s) \,. \labbel{eq:newbasissun1}
\end{align}

Note that in this case the first of the two masters in~\eqref{eq:newbasissun1}
has only a single pole in $(d-4)$ due to the exactness of relation~\eqref{eq:relsun1d4}.
As second master integral we can choose any of the two and here we performed
simply a random choice picking $\I_1(d;s)$. Choosing $\I_2(d;s)$ would indeed 
lead to  equivalent results. 
Deriving the differential equations for the new basis we find
\begin{align}
\frac{d\, \J_1}{d\,s} &= \left[ \frac{2}{s-m^2} - \frac{1}{s} 
+ (d-4) \left( \frac{2}{s-m^2} - \frac{3}{2\,s} \right)\right]\,\J_1 \nonumber \\
&+ (d-4) \left[ \frac{3}{2(s-m^2)} -\frac{1}{s} 
+ \frac{3}{2}(d-4) \left( \frac{1}{s-m^2} - \frac{1}{s}\right)\right]\, \J_2 \nonumber \\
\frac{d\,\J_2}{d\,s} &=  \frac{1}{s}\,\J_1 + \frac{(d-4)}{s}\,\J_2\,. \labbel{eq:deqsun1dec}
\end{align}
Equations~\eqref{eq:deqsun1dec} confirm the discussion in Section~\ref{sec:IBPsN} 
and can therefore be seen as of the main result of this paper.
Let us have a closer look at these two equations and compare them to~\eqref{eq:deqsun1}.
We note immediately that the equations are not in canonical form. On the other hand, the matrix of
the system does become \textsl{triangular} in the limit $d\to4$, and in particular the master $\J_2$
appears in the differential equation
for integral $\J_1$ multiplied by an explicit factor $(d-4)$,
as predicted in~\eqref{eq:gensuppr}.
For any practical purposes this is enough, since it means that one can expand the differential equations as Laurent
series in $(d-4)$ and, order by order, first solve the differential equation for $\J_1$ by simple
quadrature, and then use this result as input for the differential equation for $\J_2$, which 
can in turn be solved by quadrature. Needless to say, this procedure can in principle be iterated up to any
order in $(d-4)$.

A comment is in order. In this simple example the relation found by studying the IBPs
in $d=4$ can be interpreted as an actual relation between the double poles of the two master integrals.
Very often the first poles of arbitrarily chosen MIs are
either constants or very simple rational functions. One might therefore naively think that, by 
evaluating explicitly the poles of a given set of MIs, one could simply look for simple relations 
among the latter. Such relations would indeed contain only simple rational functions.
It is well known, though, that in several cases the poles of a master integral can be represented
entirely through its sub-topologies. If this is the case, a relation between the poles of the masters would be
useless, as it would not bring any new information as far as the master integrals are concerned.
In order to achieve a decoupling one must therefore use a relation which is contained in
the IBPs and as such represents an effective \textsl{degeneracy} of the master integrals in the limit $d \to 2\,n$,
with $n \in \N$.

\subsubsection{Simplification of the differential equations in $d=2$}\labbel{sec:Tri2}
It is interesting to see what happens
by repeating the same analysis for the graph~\eqref{eq:sunr1} in $d=2$ instead of $d=4$.  Again, an easy
analysis of the two master integrals~\eqref{eq:missun1} shows that they both develop
a double pole in $d=2$, which in this case is of IR origin
\begin{align}
&\I_1(d;s) = \frac{1}{(d-2)^2}\,\I_1^{(-2)}(2;s) + \frac{1}{(d-2)}\,\I_1^{(-1)}(2;s) + \mathcal{O}(1) \\
&\I_2(d;s) = \frac{1}{(d-2)^2}\,\I_2^{(-2)}(2;s) + \frac{1}{(d-2)}\,\I_2^{(-1)}(2;s) + \mathcal{O}(1)\,.
\end{align}
As before, one can easily see that also in this case all integrals of the form~\eqref{eq:sunr1} can develop
\textsl{at most} a double pole in $(d-2)$. We can proceed and generate the IBPs for generic $d$
and then expand them as Laurent series, this time in $(d-2)$, starting from $1/(d-2)^2$. We are then left
with a series of chained systems of IBPs, each for a different order in $(d-2)$. As in the previous case,
we can now focus on solving the first system, corresponding to the double pole. This again is equivalent to 
considering the original system of IBPs and simply fixing $d=2$.
Upon doing this one immediately sees that once more the two MIs become linearly dependent
\begin{align}
\I_2^{(-2)}(2;s) = \frac{1}{s-m^2}\I_1^{(-2)}(2;s)\,.  \labbel{eq:relsun1d2}
\end{align}
As for the previous case, relation~\eqref{eq:relsun1d2} must come from a 
corresponding $d$-dimensional relation. Indeed, if one solves the IBPs in $d$ dimensions
one finds, among the others, the following relation
\begin{align*}
I(d; 1,1,1,1,0) =
\left(\frac{1}{d-2}\right)\, \frac{m^2}{3}  \left(\,\left[ (2-d)\frac{s}{m^2} + (3-d) \right] I_1(d;s) 
- (s-m^2)I_2(d;s) \right)\,.\labbel{eq:ibpd2}
\end{align*}
It is clear that in the limit $d \to 2$ Eq.~\eqref{eq:ibpd2} generates instead Eq.~\eqref{eq:relsun1d2}.
Proceeding as above, we can choose as new basis of MIs
\begin{align}
&\J_1(d;s) = \I_1(d;s) - (s - m^2) \I_2(d;s)\,, \qquad
\J_2(d;s) = \I_1(d;s) \,.
\end{align}
Deriving the differential equations for $\J_1$ and $\J_2$ one finds
immediately 

\begin{align}
&\frac{d \J_1}{d s} = 
(d-2)  \left[ \frac{2}{s-m^2} -  \frac{3}{2\,s} \right] \J_1 
+ (d-2) \left[  \frac{3(d-2)}{2\,s} - \frac{2}{s-m^2}\right]  \J_2 
 \nonumber \\
&\frac{d \J_2}{d s} = 
\left[ \frac{1}{s-m^2} - \frac{1}{s} \right] \J_1+
\left[ \frac{(d-2)}{s} - \frac{1}{s-m^2} \right] \,\J_2\,.
\end{align}
Again, as expected from the arguments of Section~\ref{sec:IBPsN}, we see that the differential
equation for $\J_1$ decouples from the one for $\J_2$ in the limit $d\to 2$,
respecting the same pattern described in equation~\eqref{eq:gensuppr}.
Once more for every practical purposes this is enough to reduce the solution
of the system of differential equations to iterated integrations by quadrature.

\subsection{A two-loop triangle with three legs off-shell} \labbel{sec:triangle} 

Let us consider now a massless 
two-loop three-point function with three legs off-shell. The problem has been widely studied
 in the literature, mainly in the context of vector boson pair 
production~\cite{Birthwright:2004kk,Chavez:2012kn,Gehrmann:2013cxs,Henn:2014lfa}, 
and it is well known that this Feynman graph can be reduced
to two independent MIs, which can be integrated in terms of MPLs only.
We define the Feynman graph as follows

\begin{align}
I(d;n_1,&n_2,n_3,n_4,n_5,n_6,n_7) = \triangleThree{q}{p_1}{p_2} \nonumber \\ 
&= \int \D^d k  \D^d l \frac{ \left(l\cdot l  \right )^{n_5} 
\left( k \cdot p_2\right)^{n_6}
\left( l \cdot p_2 \right)^{n_7}}
{
\left( k^2 \right)^{n_1} 
\left( (k-l)^2 \right)^{n_2} 
\left( (l-p_1)^2 \right )^{n_3} 
\left( (k-p_1-p_2)^2 \right )^{n_4} 
}\,, \labbel{trian1}
\end{align}
where $p_1^2 = m_1^2$, $p_2^2=m_2^2$ and $q^2 = (p_1+p_2)^2 = s$.

We used Reduze 2 in order to reduce this graph to two independent MIs
\begin{equation}
\I_1(d;s,m_1^2,m_2^2) = I(d;1,1,1,1,0,0,0)\,,\quad \I_2(d;s,m_1^2,m_2^2) = I(d;2,1,1,1,0,0,0)\,.
\end{equation}

We can then proceed and derive the differential equations for these two MIs.
As always we neglect the sub-topologies throughout. 
In this particular case the latter are simple
two-loop corrections to massless two-point functions which have been known analytically for a very long time.

The homogeneous part of the differential equations in the momentum transfer $s$ reads
\begin{align}
& P(s,m_1^2,m_2^2)\, \frac{d\,\I_1}{d\,s} = \frac{(d-4)(m_1^2-m_2^2)^2 
+ \left( (3d-8)m_1^2 - 3(d-4)m_2^2\right)s + 2(d-4)s^2}{2s}\, \I_1 + 2\,s\,m_1^2\,\I_2 \\
& P(s,m_1^2,m_2^2)\,\frac{d\,\I_2}{d\,s} = \frac{(10-3\,d)\left( (d-3)(m_1^2-m_2^2) + (2d-7)\,s \right)}{2s} \,\I_1
\nonumber \\
&\qquad \qquad \qquad \quad \;\;\, + \frac{(d-6) (m_1^2-m_2^2)^2 +  \left( (22-5d)m_1^2 + (d+2)m_2^2 \right)\,s
-2(d-2)\,s^2 }{2s}
\, \I_2\,, \labbel{eq:deqtri} \end{align}
where we defined the polynomial
\begin{equation}
P(s,m_1^2,m_2) = m_1^4+(s-m_2^2)^2-2 m_1^2(s+m_2^2)\,.
\end{equation}

The equations are coupled in the limit $d \to 4$. 
Again, as for the sunrise studied in Section~\ref{sec:sun1}, the integrals can develop at most a double pole in $(d-4)$.
Instead of performing a complete Laurent expansion of the IBPs, 
we generate them and then fix explicitly $d=4$ before solving them.
This is enough to check whether the two MIs degenerate in this limit.
By solving the IBPs one finds that this is precisely the case and the following relation is
extracted
\begin{align}
\I_1(d;s,m_1^2,m_2^2) +s\,\I_2(d;s,m_1^2,m_2^2) = \mathcal{O}(d-4)\,, \labbel{eq:trid4}
\end{align}
where again we used the notation introduced above, indicating that the combination~\eqref{eq:trid4}
becomes zero if we fix $d=4$ in the IBPs.
Of course, also in this case, if we had expanded the IBPs as Laurent series starting from the 
double pole, relation~\eqref{eq:trid4} could have been interpreted as a relation between the double
poles of the two master integrals
\begin{align}
\I_1^{(-2)}(4;s,m_1^2,m_2^2) +s\,\I_2^{(-2)}(d;s,m_1^2,m_2^2) = 0\,.
\end{align}
Note, nevertheless, that this time the relation is \textsl{not exact}, 
differently from~\eqref{eq:relsun1d4}, since
the sub-topologies might in general contribute modifying~\eqref{eq:trid4}. 
Eq.~\eqref{eq:trid4}
 is anyway sufficient to decouple the homogeneous part of the differential equations.
We proceed as above and define the new basis
\begin{align}
\J_1(d;s,m_1^2,m_2^2)=\I_1(d;s,m_1^2,m_2^2) +s\,\I_2(d;s,m_1^2,m_2^2) \,, \quad
\J_2(d;s,m_1^2,m_2^2)=\I_1(d;s,m_1^2,m_2^2)\,. \labbel{eq:basistri2}
\end{align}
Deriving the differential equations satisfied by~\eqref{eq:basistri2} we find

\begin{align}
& P(s,m_1^2,m_2^2)\, \frac{d \J_1}{d\,s} = 
\frac{ (d-4)(m_1^2-m_2^2)^2 + \left( (22-5d)m_1^2 + (d-2)m_2^2\right)s - 2(d-3)s^2 }{2s}\,\J_1
\nonumber \\
&\qquad \qquad \qquad \quad \;\;\, 
-\frac{(d-4)\,\left( 3(d-5) m_1^2 + (11-3d)m_2^2 - 3(7-2d)s\right)}{2}\,\J_2 \nonumber \\
& P(s,m_1^2,m_2^2)\, \frac{d \J_2}{d\,s} = 2 \, m_1^2\,\J_1 
+ \frac{(d-4)(s+m_1^2-m_2^2)(2\,s+m_1^2-m_2^2)}{2\,s}\,\J_2\,.
\end{align}

Again, as expected, we see that the differential equation for $\J_1$ contains an explicit factor $(d-4)$
multiplying the second integral $\J_2$. The result is consistent with the general structure
described in Section~\ref{sec:IBPsN}.
Once more we want to stress that, as expected, in this case all MIs can be integrated in terms
of MPLs only.

\subsection{The two-loop massive sunrise} \labbel{sec:sun2}
In the previous sections we considered two simple examples of  $2 \times 2$ systems where 
the two equations could be decoupled in the limit
 $d \to 4$, such that the problem could always be reduced to integrations
 by quadrature. As we already discussed this is not always possible and the first 
known case where at least two
differential equations remain coupled is the two-loop massive sunrise.
The two-loop massive sunrise graph is defined as follows

\begin{align}
I(d;n_1,n_2,n_3,n_4,n_5) &= \Sunrisetwo{p} \nonumber \\
&= \int \D^d k  \D^d l\, \frac{(k \cdot p)^{n_4}(l \cdot p)^{n_5}}
{\left( k^2-m_1^2 \right)^{n_1} \left( l^2 -m_2^2\right )^{n_2} \left((k-l+p)^2-m_3^2 \right)^{n_3}}\,. \labbel{eq:sunr3}
\end{align}
This integral has been studied widely in the literature and in particular a lot of attention has been devoted
to the differential equations that it fulfils. 
In the general case where all three masses assume different values, a normal reduction 
through IBPs shows that all integrals can be expressed as linear combinations
of 4 independent MIs, which can be chosen to be
\begin{align}
&\I_1(d;s) = I(d;1,1,1,0,0)\,, \quad \I_2(d;s) = I(d;2,1,1,0,0)\,, \nonumber \\
&\I_3(d;s) = I(d;1,2,1,0,0)\,, \quad \I_4(d;s) = I(d;1,1,2,0,0)\,. \labbel{eq:missunr3}
\end{align}

In~\cite{Caffo:1998du} it was shown that these integrals fulfil a coupled system of 4 linear first order
differential equations in $d$ dimensions. The system remains coupled in the limits $d \to 2\,n$, where
$n \in \mathbb{N}$. It was lately shown in~\cite{MullerStach:2012mp}, 
using algebraic geometry methods (and as such \textsl{a priori} orthogonal to the IBPs), that the scalar integral
$\I_1(d;s)$ satisfies a second-order Picard-Fuchs differential equation in $d=2$. 
This suggested the
possibility of finding a proper change of basis of MIs, in the sense of the IBPs, 
such that two of the four differential equations satisfied by the latter
 would decouple in the limit $d \to 2$. Since the four MIs in~\eqref{eq:missunr3} are \textsl{finite}
in $d=2$, it appeared natural to try and obtain the decoupling of the differential equations
by finding \textsl{new relations} among the MIs, valid strictly only for $d=2$.
Such relations can be found using the so-called \textsl{Schouten Identities}
introduced in~\cite{Remiddi:2013joa}.
In that reference the Schouten identities are introduced and the case of the two-loop sunrise with
different masses is worked out in detail. It is shown that, as expected, in $d=2$ only two master
integrals are linearly independent. This allowed to recover the second order differential equation
found in reference~\cite{MullerStach:2012mp} in a completely independent manner.
In this section we will show that those results can be even more easily re-obtained 
using the methods described in this paper, and namely by solving the
IBPs for the massive sunrise in $d=2$. The Schouten identities can be imagined
as a tool for extracting this piece of information from the IBPs and are, in this respect,
equivalent to the study of the IBPs in fixed number of dimensions.
We will show another example of this equivalence in Appendix~\ref{App:Sch}.

Since the algebra in this case is rather heavy due to the large number of scales, 
we will only report the result
of the solution of the IBPs in $d=2$, referring to~\cite{Remiddi:2013joa} for their use to
simplify the system of differential equations. As already discussed above, solving the system with $d=2$
is in general easier and, as expected, two of the four MIs degenerate, leaving only
two linearly independent MIs. By choosing
as MIs $\I_1(2;s)$ and $\I_2(2;s)$, we find the following additional relations 
(as everywhere else we neglect the sub-topologies for simplicity)

\begin{align}
 m_2^2\,&P(s,m_1^2,m_2^2,m_3^2)\,\I_3(2;s) = 
(m_1^2 - m_2^2)  (m_1^2 + m_2^2 - m_3^2 - s) \, \I_1(2;s) 
\nonumber \\ &+ 
  m_1^2 \left( m_1^4 - 3 m_2^4 + 
     2 m_1^2 (m_2^2 - m_3^2 - s) + (m_3^2 - s)^2 + 
     2 m_2^2 (m_3^2 + s) \right) \I_2(2;s)  \nonumber \\ & \nonumber \\
  m_3^2\,&P(s,m_1^2,m_2^2,m_3^2)\,\I_4(2;s) = 
  (m_1^2 - m_3^2)  (m_1^2 - m_2^2 + m_3^2 - s) \I_1(2;s)  
  \nonumber \\ &+ 
  m_1^2 \left( m_1^4 + m_2^4 - 3 m_3^4 + 2 m_2^2 (m_3^2 - s) + 2 m_3^2 s + 
     s^2 - 2 m_1^2 (m_2^2 - m_3^2 + s) \right) \I_2(2;s)\,,
     \labbel{eq:relsunr3}
      \end{align}
where we defined the polynomial
$$P(s,m_1^2,m_2^2,m_3^2) = (-3 m_1^4 + m_2^4 + (m_3^2 - s)^2 - 2 m_2^2 (m_3^2 + s) + 
    2 m_1^2 (m_2^2 + m_3^2 + s)).$$
    
As we discussed in Section~\ref{sec:IBPsN}, we expect such relations to come by $d$-dimensional
IBPs with an overall factor $1/(d-2)$. Indeed by studying the reduction to MIs in $d$ dimensions
it is easy to find the following two relations

\begin{align}
\mathcal{O}(d-2) = \frac{1 }{3} & \Big\{ \left[ (d-3) (2 m_1^2 - m_2^2 - m_3^2) - (d - 2) s\right] \, \I_1(d;s)\,
+2\,m_1^2(s-m_1^2) \I_2(d;s) \nonumber \\
&+ m_2^2 (-3 m_1^2 + m_2^2 + 3 m_3^2 - s) \I_3(d;s) + m_3^2 (-3 m_1^2 + 3 m_2^2 + m_3^2 - s)\,\I_4(d;s)
\Big\}\nonumber \\&\nonumber \\
\mathcal{O}(d-2) = \frac{1 }{3} & \Big\{ \left[ (d-3) (m_1^2 - 2m_2^2 + m_3^2) - (d - 2) s\right] \, \I_1(d;s)\,
-2\,m_2^2(s-m_2^2) \I_3(d;s) \nonumber \\
&+ m_1^2 (-m_1^2 + 3m_2^2 - 3 m_3^2 + s) \I_2(d;s) + m_3^2 (-3 m_1^2 + 3 m_2^2 - m_3^2 + s)\,\I_4(d;s)
\Big\}\,. \labbel{eq:relsunr3ibps}
\end{align}
Relations~\eqref{eq:relsunr3ibps} can be compared with the corresponding formulas (3.14) and (3.15) 
of~\cite{Remiddi:2013joa}. It is easy to see that they are identical in the limit $d \to 2$,
the only difference being the absence of the terms coming from the sub-topologies,
which we are neglecting here.
These two relations (and in particular their limiting value as $d \to 2$) 
can be used, as described in Section~\ref{sec:IBPsN}, in order to decouple two of the four differential
equations of the two-loop massive sunrise graph, by choosing as new basis of master integrals
\begin{align}
&\J_1(d;s) = \I_1(d;s)\,,\quad \J_2(d;s) = \I_2(d;s)\nonumber \\&\nonumber \\
&\J_3(d;s) = - (2 m_1^2 - m_2^2 - m_3^2) \, \I_1(d;s)\,
+2\,m_1^2(s-m_1^2) \I_2(d;s) \nonumber \\
&+ m_2^2 (-3 m_1^2 + m_2^2 + 3 m_3^2 - s) \I_3(d;s) + m_3^2 (-3 m_1^2 + 3 m_2^2 + m_3^2 - s)\,\I_4(d;s)
\nonumber \\&\nonumber\\
&\J_4(d;s) = - (m_1^2 - 2m_2^2 + m_3^2) \, \I_1(d;s)\,
-2\,m_2^2(s-m_2^2) \I_3(d;s) \nonumber \\
&+ m_1^2 (-m_1^2 + 3m_2^2 - 3 m_3^2 + s) \I_2(d;s) + m_3^2 (-3 m_1^2 + 3 m_2^2 - m_3^2 + s)\,\I_4(d;s)\,.
\end{align}
We do not give the explicit form of the differential equations, referring to~\cite{Remiddi:2013joa}
for further details. In comparing, note that the basis presented here differs from the one 
in~\cite{Remiddi:2013joa} by the absence of
sub-topologies and by orders $\mathcal{O}(d-2)$. 
Furthermore we want to stress, in relation to the discussion
in Section~\ref{sec:IBPsN}, that using this basis produces an overall factor $(d-2)$ in front of the two
differential equations for $\J_3(d;x)$ and $\J_4(d;x)$. This implies that
one has, at every order in $(d-2)$, only one second order differential equation (needed to solve the block 
of $\J_1(d;x)$ and $\J_2(d;x)$), plus two integrations
by quadrature (required to determine $\J_3(d;x)$ and $\J_4(d;x)$).

\subsection{A two-loop non-planar crossed vertex} \labbel{sec:crossed}
As a further application, let us consider a two-loop non-planar crossed vertex with two massive propagators.
This graph is topologically completely unrelated to the two-loop sunrise and was studied thoroughly
 in~\cite{Aglietti:2007as}. There it was shown
that it can be reduced to three MIs, which would therefore be expected to satisfy a system
of three coupled differential equations. In~\cite{Aglietti:2007as} a basis of MIs was found
such that one of the three differential equations decouples from the other two in the limit
$d \to 4$. This reduced effectively the problem to that of solving, for every order in $(d-4)$, a second order
differential equation, plus an integration by quadrature for the third MI.
In this section we would like to study this Feynman graph with our method and
show that the decoupling found in~\cite{Aglietti:2007as} comes as well from a degeneracy of
the master integrals in $d=4$ which can be read off directly from the IBPs.
Following~\cite{Aglietti:2007as} we define the Feynman graph as follows

\begin{align}
I(&d;n_1,n_2,n_3,n_4,n_5,n_6,n_7) = \trianglecross{q}{p_1}{p_2} \nonumber \\ 
&=\int \D^d k\, \D^d l\, 
\frac{(\,k \cdot p_2)^{n_7}}
{(k^2-m^2)^{n_1} (l^2-m^2)^{n_2} \left((k-p_1)^2\right)^{n_3} \left((l-p_2)^2\right)^{n_4}
\left((k-l-p_1)^2\right)^{n_5} \left((k-l+p_2)^2\right)^{n_6}}\,,
\end{align}
with $p_1^2 = p_2^2 = 0$ and $(p_1+p_2)^2 = s$.
It is easy to verify that this topology can be reduced to 3 MIs, for example
\begin{align}
\I_1(d;s) = I(d;1,1,1,1,1,1,0)\,, \quad \I_2(d;s) = I(d;2,1,1,1,1,1,0)\,, \quad \I_3(d;s) = I(d;1,1,2,1,1,1,0)\,.
\labbel{eq:miscrossed}
\end{align}

Let us derive the differential equations in the momentum transfer $s$, neglecting as everywhere else all
sub-topologies. We get

\begin{align}
\frac{d\,\I_1}{d\,s} &= \frac{(d-6)}{s}\I_1 - \frac{2\,m^2}{s}\I_2\,, \nonumber \\ & \nonumber \\
\frac{d\,\I_2}{d\,s} &=  \frac{(d-5)(2d-9)(s-4m^2)}{s(s-m^2)(s+8m^2)}\,\I_1
+ \frac{14 (d-4) m^4 - (5 d-13) m^2 s - 2 s^2}{s(s-m^2)(s+8m^2)}\,\I_2 \nonumber \\
&
+ \frac{2 \,(d-4) \,m^2}{s(s+8\,m^2)}\,\I_3\,, \nonumber \\ & \nonumber \\
\frac{d\,\I_3}{d\,s} &=  \frac{2(d-5)(2d-9)(s+2m^2)}{s(s-m^2)(s+8m^2)}\,\I_1
+ \frac{2 m^2\left( (24 - 5 d) m^2 + (21 - 4 d) s \right) }{s(s-m^2)(s+8m^2)}\,\I_2 \nonumber \\
&
- \frac{2 (3 d-8) m^2 + (d-3) s}{s(s+8\,m^2)}\,\I_3 \,.
\labbel{eq:deqcrossed}
\end{align}

Looking at these equations we see immediately that $\I_3$ is already decoupled from
the other two in the limit $d \to 4$. This means that, with this basis, the problem is
reduced to that of solving, at every order in $(d-4)$, a coupled system for $\I_1$ and $\I_2$.
With the explicit solution for the latter, one can then obtain $\I_3$ integrating its differential equation by quadrature.
It would be then interesting to know whether this decoupling is also due to a degeneracy of the MIs
in $d=4$. Moreover it would be even more interesting to verify
whether a new basis could be found, for which the differential equations become
completely triangular as $d \to 4$, reducing even further the complexity of the problem.

Following the recipe described above, we can try and solve the IBPs for this Feynman graph for $d=4$.
A word of caution is required here. The three MIs selected above have Laurent expansions in 
$(d-4)$ which start at different orders, in particular one can easily find 
(for example using sector decomposition~\cite{Borowka:2015mxa}) that the first two masters
are finite, while the third develops a cubic pole
\begin{align}
\I_1(d\to4;s) = \mathcal{O}(1)\,,\quad \I_2(d\to4;s) = \mathcal{O}(1)\,, \quad
\I_3(d\to4;s) = \mathcal{O} \left( \frac{1}{(d-4)^3} \right)\,. \label{eq:polescrossed}
\end{align}

Nevertheless this poses no practical obstacle to the applicability of the method presented in this paper.
As we already discussed in general, by expanding the system of IBPs in Laurent series around $d=4$, 
we will get, at every order in $(d-4)$, an independent system of differential equations whose homogeneous
part (i.e. the one containing the order of the MIs under consideration) has always the same form, 
while the non-homogeneous part will of course change and, in particular, 
depend on the previous orders of the expansion (and on the sub-topologies, which we neglect). 
What we are interested in is, indeed, only the homogeneous part of this system. 
Fixing $d=4$ is therefore enough in order to determine
whether, for any order of the expansion, the MIs become linearly dependent. Upon doing this we find only
one relation between the three masters which reads
\begin{equation}
\I_1(d;s) + (5\,m^2+s)\I_2(d;s)  + 3\,m^2\,\I_3(d;s) = \mathcal{O}(d-4)\,, \labbel{eq:relcrossed1}
\end{equation}
where, as always, we mean that this combination becomes zero when we set $d=4$ in the IBPs.

Equivalently, one can also proceed in a more formal way, expanding all IBPs in Laurent series
starting from the triple pole up to the finite piece, and supplementing the piece of information 
on the highest poles of the MIs~\eqref{eq:polescrossed}. Upon doing this, one obtains four chained systems
of IBPs (one for every oder in $(d-4)$), 
which can be solved bottom-up starting from the one corresponding to the highest pole.
Since the first two masters are finite, the first three systems give no information on the latter, while
the fourth system (corresponding to the finite piece of the MIs) produces the relation
\begin{equation}
\I_1^{(0)}(4;s) + (5\,m^2+s)\I_2^{(0)}(4;s)  + 3\,m^2\,\I_3^{(0)}(4;s)  + \frac{2}{m^2} I^{(-1)}(4;1,1,1,1,1,1,1)=0\,.
\labbel{eq:relcrossed2}
\end{equation}
Indeed relations~\eqref{eq:relcrossed1} and~\eqref{eq:relcrossed2} are identical up
to the presence of the previous order in the expansion of the integral $I(d;1,1,1,1,1,1,1)$. If we had
solved the IBPs in $d$ dimensions, this integral would have been of course expressed in terms of the 
three masters~\eqref{eq:deqcrossed}. Solving the system in $d=4$ instead does not allow to express
this integral in terms of the other three, but this comes with no surprise and can be very well understood
in terms of the degeneracy of the system of IBPs in this limit.

By studying explicitly the integral $I(d;1,1,1,1,1,1,1)$ it is easy to see that it is also 
finite in $d=4$, namely
$$I(d\to4;1,1,1,1,1,1,1) = \mathcal{O}(1)\,, \quad \longrightarrow \quad I^{(-1)}(4;1,1,1,1,1,1,1) = 0\,.$$
With this piece of information one recovers again relation~\eqref{eq:relcrossed1}, 
which was found by simply solving the system of IBPs in $d=4$.  
Since only one relation has been found, which moreover involves all three masters $\I_1$, $\I_2$, and $\I_3$, 
we have no way to decouple more than one MIs from the system.
What we mean here is that, since in system~\eqref{eq:deqcrossed} only the differential equations for $\I_1$
and $\I_2$ are coupled, if we had found one relation but involving only $\I_1$ and $\I_2$, we could
have used it to decouple this block of the system. An example of this is given in Section~\ref{sec:triangle2}.
This is not the case and we therefore expect
the minimal number of coupled integrals in $d=4$ to be two, giving rise to a second order differential equation,
as for the case of the two-loop massive sunrise, see Section~\ref{sec:sun2}.

As an exercise, we can try to change basis also in this case exploiting the piece of information
found in~\eqref{eq:relcrossed1}. We expect to end up with a new system of differential equations,
where nevertheless again two out of three equations are coupled as $d \to 4$ 
(and as such practically equivalent to~\eqref{eq:deqcrossed}),
showing that the system cannot be further simplified. Let us introduce the new basis
\begin{align}
\J_1(d;s) = \I_1(d;s)\,,\quad \J_2(d;s) = \I_2(d;s)\,, \quad 
\J_3(d;s) = \I_1(d;s) + (5\,m^2+s)\I_2(d;s)  + 3\,m^2\,\I_3(d;s)\,. 
\end{align}
Deriving the differential equations and neglecting all sub-topologies we get
\begin{align}
\frac{d\, \J_1}{d\,s} &=  \frac{(d-6)}{s}\J_1 - \frac{2\,m^2}{s}\J_2 \nonumber \\ & \nonumber \\
\frac{d\, \J_2}{d\,s} &=  \frac{2(d-4)(s-m^2)+3(d-5)(2d-9)(s-4m^2)}{3s(s-m^2)(s+8m^2)}\,\J_1 
 \nonumber \\ 
&+ \frac{52(d-4)m^4-(23d-71)m^2\,s-2(d-1)s^2}{3\,s(s-m^2)(s+8m^2)}\,\J_2
+ \frac{2 \,(d-4) \,m^2}{s(s+8\,m^2)}\,\J_3 \nonumber \\ & \nonumber \\
\frac{d\, \J_3}{d\,s} &=  \frac{(d-4) (6 d-29)}{9 m^2 s} \J_1 + \frac{(d-4)(s-10m^2)}{9 m^2 s} \J_2
- \frac{(d-1) \J_3}{3\, s}\,,  \nonumber \\ \labbel{eq:deqcrossed2}
 \end{align}
which is again a system of three differential equation, two of which remain coupled in the limit $d \to 4$,
giving rise to a second order differential equation for one of the two coupled 
masters.\footnote{In~\cite{Aglietti:2007as} it was shown that the homogeneous
part of the second order differential equation
satisfied by the scalar master integral $\I_1(d;s)$ is equivalent to that of the two-loop massive
sunrise with equal masses.} We note that the new system~\eqref{eq:deqcrossed2}, compared with 
the previous one~\eqref{eq:deqcrossed}, has a slightly different structure. As for the previous cases
that we analysed, once we switch to the new basis defined through the IBPs degeneracy~\eqref{eq:relcrossed1},
the differential equation for the new master $\J_3$ develops an explicit factor $(d-4)$ in front of 
the other two masters $\J_1$ and $\J_2$, as predicted in equation~\eqref{eq:gensuppr}.

\subsection{A two-loop massive triangle with three master integrals} \labbel{sec:triangle2}
Before moving to a three-loop example, let us try and see what happens in a case similar to
the one studied above, i.e. a Feynman graph reduced to three master integrals, but where
the system of differential equations can be completely triangularised as $d \to 4$.
Let us consider the following two-loop massive triangle

\begin{align}
I(d;n_1,&n_2,n_3,n_4,n_5,n_6,n_7) = \triangleMass{P}{p_1}{p_2} \nonumber \\
&= \int \D^d k  \D^d l \frac{ \left(k\cdot p_1  \right )^{n_5} 
\left( l \cdot p_1\right)^{n_6}
\left( l \cdot q \right)^{n_7}}
{
\left( l^2 - m^2\right)^{n_1} 
\left( (k-l)^2 \right)^{n_2} 
\left( (k-p_1)^2 -m^2\right )^{n_3} 
\left( (k-p_1-p_2)^2-m^2 \right )^{n_4} 
}\,, \labbel{trian2}
\end{align}
with two legs off-shell, namely $P^2 = (p_1+p_2) = s$, $p_1^2 = 0$, $p_2^2 = q^2$.
This graph has been studied in the context of the QCD corrections to $H \to Z\gamma$ in~\cite{Bonciani:2015eua,Gehrmann:2015dua}. Similarly to our previous example it
is reduced to three master integrals, such that we are dealing  with a system of
three differential equations. We start from an arbitrarily chosen basis of master integrals
\begin{align}
\I_1(d;s,q^2) = I(d;1,1,&1,1,0,0,0)\,, \quad \I_2(d;s,q^2) = I(d;1,1,2,1,0,0,0)\nonumber \\
&\I_3(d;s,q^2) = I(d;1,1,1,2,0,0,0)\,.
\end{align}
The masters depend on three variables $s, q^2$ and $m^2$, and therefore on 
two independent ratios. For simplicity we will consider
only the differential equations in $s$, while all considerations done here work identically
for the differential equations in the other variables.
In order to simplify as much as possible the formulas we write explicitly only the order zero
of the homogeneous differential equations in $(d-4)$, which is also the bulk which we need to
simplify. The equations read

\begin{align}
&\frac{\partial}{\partial s} \I_1(d;s) = 
\frac{1}{(q^2-s)} \I_1(d;s) + \frac{2\,m^2}{s}\I_2(d;s) + \frac{s\,q^2 - 2 m^2(s+q^2)}{s(q^2-s)}\I_3(d;s) 
+ \mathcal{O}(d-4) \nonumber \\
&\frac{\partial}{\partial s} \I_2(d;s) = \frac{m^2}{s(q^2-s)-m^2q^2}\,\I_2(d;s)
+ \frac{q^2 s - m^2(2 q^2+s)}{(q^2-s)(s(q^2-s)-m^2q^2)}\,\I_3(d;s)  
+ \mathcal{O}(d-4) \nonumber \\
&\frac{\partial}{\partial s} \I_3(d;s) = \frac{m^2(q^2-s)}{s\,(s(q^2-s)-m^2q^2)}\,\I_2(d;s)
+ \frac{m^2(s^2+s \,q^2 - q^4)}{s(q^2-s)(s(q^2-s)-m^2q^2)}\,\I_3(d;s) 
+ \mathcal{O}(d-4) \,,
\end{align}
where the dependence from $q^2$ is left as implicit in the master integrals for ease of notation.
One can immediately see that only two of the differential equations are coupled.
One should in principle first solve the $2 \times 2$ coupled system for 
$\I_2(d;s)$ and $\I_3(d;s)$, and then, with the latter as an input, one could attempt to solve the differential
equation for $\I_1(d;s)$ by quadrature. 

Let us try now and study the IBPs in the limit $d \to 4$. By solving them as discussed in the previous sections
one sees that  the master integrals $\I_2(d;s)$, $\I_3(d;s)$ become linearly dependent in this limit and one
finds the relation
\begin{align}
 (q^2 - s )(s-2\,m^2) \I_2(d;s) + \left[ s(s+2 m^2) - 2 \,q^2\,(s-m^2) \right] \I_3(d;s) = \mathcal{O}(d-4) \,.
\labbel{eq:reltri2}
\end{align}
As for the case of the non-planar triangle studied in Section~\ref{sec:crossed}, we find only one relation, 
while we have three master integrals. One of the three masters nevertheless is already
decoupled, and moreover relation~\eqref{eq:reltri2} involves only 
$\I_2(d;s)$ and $\I_3(d;s)$, which are precisely the
two coupled integrals. We expect this therefore to be enough to decouple the system.
We define the new basis

\begin{align}
&\J_1(d;s) = \I_1(d;s)\,,\qquad \J_2(d;s) = \I_2(d;s) \nonumber \\
& \J_3(d;s) = \frac{(q^2 - s )(s-2\,m^2)}{m^4} \I_2(d;s) + \frac{ s(s+2 m^2) - 2 \,q^2\,(s-m^2) }{m^4}\I_3(d;s)\,,
\labbel{eq:newbasistri2}
\end{align}
where the $1/m^4$ has been added for dimensional reasons. Deriving the differential equations for this new basis,
and keeping again only the first order in $(d-4)$ we find

\begin{align}
\frac{\partial}{\partial s} \J_1(d;s) &= \frac{1}{q^2 -s} \J_1(s;d) + \frac{s(q^2-4 m^2)}{2\,q^2(s-m^2)-s(s+m^2)}\J_2(d;s)
\nonumber \\
&+ \frac{( 2 \,m^2 (q^2+s) - s \, q^2 ) \, m^4  }{s(q^2-s)(2q^2(s-m^2)-s(s+m^2))} \J_3(d;s) 
+ \mathcal{O}(d-4) \nonumber \\ & \nonumber \\
\frac{\partial}{\partial s} \J_2(d;s) &= 
\frac{q^2(2 \, m^4 + (s-2m^2)s )}{(q^2(s-m^2) -s^2)(2\, q^2(s-m^2)-s(s+2 m^2))}\,  \J_2(d;s) \nonumber \\
&+ \frac{(s \, m^2 - q^2(s-m^2))\,m^4}{(q^2-s) (q^2(s-m^2) -s^2)(2\, q^2(s-m^2)-s(s+2 m^2))}\, \J_3(d;s)
 + \mathcal{O}(d-4) \nonumber \\ & \nonumber \\
 \frac{\partial}{\partial s} \J_3(d;s) &= \frac{q^2 - 2 s}{s (q^2-s)}\, \J_3(d;s) + \mathcal{O}(d-4)\,.
\end{align}

As expected the system of differential equations becomes triangular and, in particular, the
equation for the new integral $\J_3(d;s)$, defined through relation~\eqref{eq:reltri2}, decouples from
$\J_2(d;s)$, following the usual pattern of equation~\eqref{eq:gensuppr}. One can then proceed, order by order in $(d-4)$, integrating by quadrature first the differential equation
for $\J_3(d;s)$, then the one for $\J_2(d;s)$ and finally the one for $\J_1(d;s)$.
As a last comment we want to stress that, if we had considered the system in $\partial/\partial q^2$,
the same change of basis~\eqref{eq:newbasistri2} would have indeed been sufficient to 
triangularise this one as well.

\subsection{The three-loop massive banana graph}\labbel{sec:ban}
As last example let us consider a more complicated three-loop graph.
We choose the three-loop massive banana graph, which is the natural three-loop 
generalisation of the two-loop massive sunrise. In the most general case
this Feynman graph depends on the momentum squared $p^2=s$ and on
four different masses $m_1$, $m_2$, $m_3$ and $m_4$
\begin{align}
I_4(d;n_1,n_2,n_3,&n_4,n_5,n_6,n_7,n_8,n_9) = \threeban{p} \nonumber \\
&=\int \D^d k_1\, \D^d k_2\, \D^d k_3\,
\frac{(k_1\cdot p)^{n_5}(k_2 \cdot p)^{n_6}(k_3 \cdot p)^{n_7}(k_1 \cdot k_2)^{n_8}(k_1 \cdot k_3)^{n_9}}
{(k_1^2-m_1^2)^{n_1}(k_2^2-m_2^2)^{n_2}(k_3^2-m_3^2)^{n_3}((k_1+k_2+k_3-p)^2-m_4^2)^{n_4}}\,,
\end{align}
where the subscript $4$ indicates that the four masses are all different.
In the two-loop case there are 4 MIs when the 3 masses have all 
different values, which in turn degenerate
to 2 MIs in the case of equal masses. 
On the other hand we saw that, irrespective of the explicit values of the internal masses, one is always left
with only two independent MIs in $d=2$~\eqref{eq:relsunr3}.
This allowed us to decouple two of the four MIs from the differential equations in the limit $d\to2$
and prove that the scalar amplitude satisfies a second order differential equation
in this limit. 

It would therefore be interesting to verify whether a similar behaviour can also be seen
in the three-loop banana graph. Since in the general case with four different masses the algebra becomes very
cumbersome, we will consider different cases of increasing complexity, namely increasing at every 
step the number of different internal masses and check how many MIs are found in $d$
dimensions and how many can be decoupled in the limit $d \to 2$. 

\subsubsection{The equal-mass case}
Let us start considering the equal-mass case. We use the following notation
$$I_1(d;n_1,n_2,n_3,n_4,n_5,n_6,n_7,n_8,n_9) = 
 I_4(d;n_1,n_2,n_3,n_4,n_5,n_6,n_7,n_8,n_9) \Big|_{m_4 = m_3 = m_2 = m_1 = m}\,,$$
where the subscript ``$1$'' indicates now that all masses have the same value.
Running a reduction to MIs with a code of choice it is easy to check that there are 3 independent MIs in $d$
dimensions which can be chosen  to be
\begin{align}
\I_1(d;s) = I_1(d;1,1,1,&1,0,0,0,0,0)\,, \quad  \I_2(d;s) = I_1(d;2,1,1,1,0,0,0,0,0)\,,\nonumber \\
& \I_3(d;s) = I_1(d;3,1,1,1,0,0,0,0,0)\,.
\end{align}

The differential equations in the momentum transfer for these three MIs read
\begin{align}
\frac{d \I_1}{d\,s} &= \frac{3d-8}{2\,s}\I_1 - \frac{4\,m^2}{s}\,\I_2 \nonumber \\ &\nonumber \\
\frac{d \I_2}{d\,s} &=  \frac{(3d-8)(2d-5)}{8\,s\,m^2}\, \I_1 
+ \frac{ (d-4)\,s - 8(2d-5)m^2 }{8\,s\,m^2} \,\I_2 - \frac{1}{2}\,\I_3
\nonumber \\  &\nonumber \\
\frac{d \I_3}{d\,s} &=  
\frac{(2d-5)(3d-8) \left(16(11d-37)m^4 + 4(32-9d)\,m^2\,s + (d-4)\,s^2 \right)}{32\,m^2\,s\,(s-4 m^2)(s-16 m^2)}\,\I_1\nonumber \\ 
& \hspace{-0.5cm} - \frac{\left[ 64 \left( 440 + (47 d-289) d \right) m^6 - 16 \left( 668 + d (62 d-409) \right) m^4 s +
 16 (d-4) (4 d-13) m^2 s^2 - (d-4)^2 s^3\right]}{32\,m^2\,s\,(s-4 m^2)(s-16 m^2)}\,\I_2 \nonumber \\
 &\hspace{-0.5cm}+ \frac{\left[ 1024 (d-4) m^8 + 192 (27 - 8 d) m^6 s + 96 (2 d-7) m^4 s^2 - 
 4 (d-4) m^2 s^3\right]}{32\,m^2\,s\,(s-4 m^2)(s-16 m^2)} \, \I_3\,,
\end{align}
and we can easily verify that, in spite of the fact that $\I_3$ does not appear in the first equation,
the system is still coupled as $d \to 2$.
Trying to solve the system of IBPs in $d=2$ shows no further degeneracy and therefore we conclude that
the system cannot be further simplified with our method. Having a system of three coupled first-order
equation means that we can rephrase it as a third-order differential equation for any of the three masters, and in
particular for the scalar amplitude $\I_1(d;s)$. The fact that the scalar amplitude fulfils a third-order
differential equation is in agreement with the findings in~\cite{MullerStach:2012mp}.
Deriving the third-order differential equation satisfied by $\I_1(d;s)$ we find
\begin{equation}
D_d^{(3)}\, \I_1(d;s) = 0\,,
\end{equation}
where the $d$-dimensional third order differential operator $D_d^{(3)}$ reads
\begin{align}
 D_d^{(3)} = &\frac{d^3}{d\,s^3} 
 + \frac{3\left( 64m^4 + 10(d-5)m^2 s - (d-4)s^2\right)}{s(s-4m^2)(s-16m^2)}\, \frac{d^2}{d\,s^2} \nonumber \\
& +\frac{(d-4)(11d-36) s^2-64(d-4)d\,m^4 - 4\left( 216+d(7d-88)\right)m^2\,s }{4\,s^2(s-4m^2)(s-16m^2)}\, \frac{d}{d\,s}
\nonumber \\
& + \frac{(3-d)(3d-8)\left( 2(d+2)m^2 + (d-4)s\right)}{4\,s^2\,(s-4m^2)(s-16m^2)}\,,
\end{align}
and all sub-topologies are neglected as always.
In the limit $d \to 2$ the differential operator simplifies to
\begin{align}
 D_2^{(3)} = &\frac{d^3}{d\,s^3} 
 + \frac{6\left( s^2 - 15 m^2 s + 32 m^4\right)}{s(s-4m^2)(s-16m^2)}\, \frac{d^2}{d\,s^2}
  +\frac{\left(7 s^2 - 68 m^2 s + 64 m^4\right) }{s^2(s-4m^2)(s-16m^2)}\, \frac{d}{d\,s}
 + \frac{1}{s^2\,(s-16m^2)}\,,
\end{align}
which is in agreement with~\cite{MullerStach:2012mp}.

\subsubsection{The case of two different masses}
Let us move now to a slightly more general case and let the masses take two different values.
There are two possible arrangements, which we call $I_2^A$ and $I_2^B$, defined
as follows
$$I_2^A(d;n_1,n_2,n_3,n_4,n_5,n_6,n_7,n_8,n_9) = 
 I_4(d;n_1,n_2,n_3,n_4,n_5,n_6,n_7,n_8,n_9) \Big|_{m_3 = m_2 = m_1=m_a, m_4=m_b}\,,$$
$$I_2^B(d;n_1,n_2,n_3,n_4,n_5,n_6,n_7,n_8,n_9) = 
 I_4(d;n_1,n_2,n_3,n_4,n_5,n_6,n_7,n_8,n_9) \Big|_{m_2 = m_1=m_a , m_4 = m_3=m_b}\,.$$
The two configurations are intrinsically different and it makes sense to look at the two cases separately.

\begin{itemize}
\item[A)] In configuration A a reduction to MIs for generic $d$ gives 5 independent MIs which can be chosen as
\begin{align}
\I_1^A(d;s) = I_2^A(d;1,1,1,&1,0,0,0,0,0)\,, \quad  \I_2^A(d;s) = I_2^A(d;2,1,1,1,0,0,0,0,0)\,,\nonumber \\
 \I_3^A(d;s) = I_2^A(d;1,1,1,&2,0,0,0,0,0)\,,\quad  \I_4^A(d;s) = I_2^A(d;3,1,1,1,0,0,0,0,0)\,,\nonumber \\
& \I_5^A(d;s) = I_2^A(d;2,2,1,1,0,0,0,0,0)\,.
\end{align}

\item[B)] In configuration B we find instead 6 independent MIs for generic $d$
\begin{align}
\I_1^B(d;s) = I_2^B(d;1,1,1,&1,0,0,0,0,0)\,, \quad  \I_2^B(d;s) = I_2^B(d;2,1,1,1,0,0,0,0,0)\,,\nonumber \\
 \I_3^B(d;s) = I_2^B(d;1,1,2,&1,0,0,0,0,0)\,,\quad  \I_4^B(d;s) = I_2^B(d;3,1,1,1,0,0,0,0,0)\,,\nonumber \\
  \I_5^B(d;s) = I_2^B(d;2,2,1,&1,0,0,0,0,0)\,, \quad \I_6^B(d;s) = I_2^B(d;2,1,2,1,0,0,0,0,0)\,.
\end{align}
\end{itemize}

A natural question at this point would be how many MIs degenerate in the two cases in the limit $d \to 2$, and
therefore what is the order of the differential equation satisfied by the scalar amplitudes $\I_1^A(d;s)$ and 
$\I_2^B(d;s)$ respectively. 
A naive expectation, based on the two-loop sunrise,
would be to see in cases $A$ and $B$, $2$ and $3$ MIs decouple respectively, such that
the problem would reduce to the solution of a third-order differential equation, as in the equal-mass case.
Unfortunately this naive expectation is not satisfied and we find that, by solving the IBPs in $d=2$, in both cases \textsl{four} MIs remain independent, corresponding in principle to a \textsl{fourth-order}
differential equation for the scalar amplitude in both mass-configurations.
On the other hand, it is interesting to see that in both configurations, in spite of the different number of MIs in
$d$ dimensions, the problem can be reduced to an equation of the same order (i.e. four) in $d=2$.

Neglecting the sub-topologies we find in configuration A the following relation which allows to express the
fifth master integral in terms of the previous four
\begin{align}
m_a^2(s-5 m_a^2 + m_b^2)\I_5^A(2;s) &= \frac{3 m_a^2 + m_b^2 -s }{12 \,m_a^2}\, \I_1^A(2;s)
+ \frac{51 m_a^4 + (m_b^2 - s)^2 - 6 m_a^2 (m_b^2 + 2 s) }{12 \,m_a^2}\, \I_2^A(2;s) \nonumber \\
&+ \frac{ m_b^2 (m_b^2 - s)}{6 \, m_a^2} \, \I_3^A(2;s)  + 
  \frac{21 m_a^4 + (m_b^2 - s)^2 - 6 m_a^2 (m_b^2 + s) }{6}\, \I_4^A(2;s)\,. \labbel{eq:relban2mA}
\end{align}

In configuration B, instead, there are two different relations, which can be used to two express 
$\I_5^B$ and $\I_6^B$ in terms of the other four MIs in $d=2$.
We do not report the explicit solution of the IBPs in $d=2$ which looks rather cumbersome.
As for the case of the two-loop sunrise, these identities originate from $d$-dimensional IBPs
which degenerate in the limit $d \to 2$ due to an overall factor $1/(d-2)$. There are many of
these relations, but only two of them are linearly independent in the limit $d \to 2$, and they read 
(keeping only the first order in $(d-2)$)
\begin{align}
\mathcal{O}(d-2) &=\left[ 2(m_a^2 + m_b^2)- s \right] \I_1(d;s) 
+  \left[ 4\,m_a^2(5 m_a^2 + 4m_b^2) - 4\,(2 m_a^2 + m_b^2) s + s^2\right]\, \I_2(d;s) 
\nonumber \\
 &+ 4 m_b^2 (2 m_a^2 + m_b^2 - s) \, \I_3(d;s) 
 + 2 m_a^2 \left[ 4(m_a^2 + m_b^2)(2 m_a^2 - s) + s^2 \right] \, \I_4(d;s)
 \nonumber \\
 &+ 4 m_a^4 \left[ 2(m_a^2 + m_b^2) -s\right]\, \I_5(d;s) + 8 m_a^2 m_b^2 (4 m_a^2-s)\,  \I_6(d;s)\,,
  \labbel{eq:relban2mB1} 
\end{align}

\begin{align}
\mathcal{O}(d-2) &= (-2 m_a^2 + 6 m_b^2 - 3 s)\,\I_1(d;s) 
+ \left[ -20 m_a^4 + 8m_a^2(7 m_b^2 -2\,s) + 3 \, s (s - 4 m_b^2)\right]\,\I_2(d;s)
\nonumber \\
&+ 12 \, m_b^2 (m_b^2-s)\,\I_3(d;s) 
+ 2 m_a^2\left[ -8 m_a^2(m_a^2 -3 m_b^2) - 4 (m_a^2 + 3 m_b^2)\,s + 3 s^2\right] \I_4(d;s)
\nonumber \\
&- 4 m_a^4 (2 m_a^2 - 6 m_b^2 + s ) \,\I_5(d;s) + 32\, m_a^2\, m_b^2 (m_b^2-s)\, \I_6(d;s)\,.
\labbel{eq:relban2mB2} 
\end{align}

We stress again that relations~\eqref{eq:relban2mA},~\eqref{eq:relban2mB1} and~\eqref{eq:relban2mB2} 
are \textsl{not exact}
since all sub-topologies have been neglected throughout. These relations can be nevertheless used
in order to derive new systems of differential equations where, for both A and B configurations, 
only $4$ equations remain coupled in the limit $d \to 2$. 
For example, in the case of configuration A we can take as new basis
$$\J_1^A(d;s) = \I_1^A(d;s)\,, \quad \J_2^A(d;s) = \I_2^A(d;s)\,, \quad\J_3^A(d;s) = \I_3^A(d;s)\,, \quad 
\J_4^A(d;s) = \I_4^A(d;s)\,,$$
plus the new master defined as
\begin{align}
\J_5^A(d;s) &=  m_a^2(s-5 m_a^2 + m_b^2)\I_5^A(2;s) - \frac{3 m_a^2 + m_b^2 -s }{12 \,m_a^2}\, \I_1^A(2;s)
\nonumber \\
&
- \frac{51 m_a^4 + (m_b^2 - s)^2 - 6 m_a^2 (m_b^2 + 2 s) }{12 \,m_a^2}\, \I_2^A(2;s) \nonumber \\
&- \frac{ m_b^2 (m_b^2 - s)}{6 \, m_a^2} \, \I_3^A(2;s)  - 
  \frac{21 m_a^4 + (m_b^2 - s)^2 - 6 m_a^2 (m_b^2 + s) }{6}\, \I_4^A(2;s)\,.
  \end{align}
Upon doing this one finds that the differential equation for the new master $\J_5^A$ assumes the form
\begin{align}
\frac{d\,\J_5}{d\,s} = (d-2) \left[ c_{51}(d;s)\J_1 + c_{52}(d;s)\J_2 + c_{53}(d;s)\J_1 + c_{54}(d;s)\J_2 \right]
+ c_{55}(d;s)\J_5\,,
\end{align}
where the functions $c_{ij}(d;s)$ are rational functions for the dimension $d$, the momentum $s$ and the two
masses, and are \textsl{finite} as $d \to 2$. This insures, thanks to the overall coefficients $(d-2)$, that
the differential equation for $\J_5$ decouples completely from the other four, as expected.

As far as configuration $B$ is concerned, in order to achieve the complete decoupling of two
out of the six equations, we can take as basis 
\begin{align}
\J_1^B(d;s) = \I_1^B(d;s)\,, \quad \J_2^B(d;s) = \I_2^B(d;s)\,, \quad\J_3^B(d;s) = \I_3^B(d;s)\,, \quad 
\J_4^B(d;s) = \I_4^B(d;s)\,,
\end{align}
together with
\begin{align}
\J_5^B(d;s) &=\left[ 2(m_a^2 + m_b^2)- s \right] \I_1(d;s) 
+  \left[ 4\,m_a^2(5 m_a^2 + 4m_b^2) - 4\,(2 m_a^2 + m_b^2) s + s^2\right]\, \I_2(d;s) 
\nonumber \\
 &+ 4 m_b^2 (2 m_a^2 + m_b^2 - s) \, \I_3(d;s) 
 + 2 m_a^2 \left[ 4(m_a^2 + m_b^2)(2 m_a^2 - s) + s^2 \right] \, \I_4(d;s)
 \nonumber \\
 &+ 4 m_a^4 \left[ 2(m_a^2 + m_b^2) -s\right]\, \I_5(d;s) + 8 m_a^2 m_b^2 (4 m_a^2-s)\,  \I_6(d;s) \,,
\end{align}
\begin{align}
\J_6^B(d;s) &=(-2 m_a^2 + 6 m_b^2 - 3 s)\,\I_1(d;s) 
+ \left[ -20 m_a^4 + 8m_a^2(7 m_b^2 -2\,s) + 3 \, s (s - 4 m_b^2)\right]\,\I_2(d;s)
\nonumber \\
&+ 12 \, m_b^2 (m_b^2-s)\,\I_3(d;s) 
+ 2 m_a^2\left[ -8 m_a^2(m_a^2 -3 m_b^2) - 4 (m_a^2 + 3 m_b^2)\,s + 3 s^2\right] \I_4(d;s)
\nonumber \\
&- 4 m_a^4 (2 m_a^2 - 6 m_b^2 + s ) \,\I_5(d;s) + 32\, m_a^2\, m_b^2 (m_b^2-s)\, \I_6(d;s)\,.
\end{align}
Using this basis one obtains a new system of differential equations, where the two equations
for $\J_5^B$ and $\J_6^B$ develop an explicit overall factor $(d-2)$, such that, at every oder
in the Laurent expansion, they can be solved trivially by quadrature. Order by order, 
once the result for the
latter is known, one is left with a system of four coupled differential equations for the remaining master
integrals.
For compactness we prefer not to give here explicitly 
the systems of differential equations.

\subsubsection{The case of three different masses}
Generalising even further we can check what happens if three out of the four masses 
are allowed to take different values.
In this case there is of course only one possibility, which we choose to be
$$I_3(d;n_1,n_2,n_3,n_4,n_5,n_6,n_7,n_8,n_9) = 
 I_4(d;n_1,n_2,n_3,n_4,n_5,n_6,n_7,n_8,n_9) \Big|_{m_4 = m_3}\,.$$

We start, as always, performing a reduction for generic $d$. The complexity increases
and we find $8$ independent MIs
\begin{align}
\I_1(d;s) = I_3(d;1,1,1,&1,0,0,0,0,0)\,, \quad  \I_2(d;s) = I_3(d;2,1,1,1,0,0,0,0,0)\,,\nonumber \\
 \I_3(d;s) = I_3(d;1,2,1,&1,0,0,0,0,0)\,,\quad  \I_4(d;s) = I_3(d;1,1,2,1,0,0,0,0,0)\,,\nonumber \\
  \I_5(d;s) = I_3(d;3,1,1,&1,0,0,0,0,0)\,, \quad \I_6(d;s) = I_3(d;2,2,1,1,0,0,0,0,0)\,,\nonumber \\
  \I_7(d;s) = I_3(d;2,1,2,&1,0,0,0,0,0)\,, \quad \I_8(d;s) = I_3(d;1,2,2,1,0,0,0,0,0)\,.
\end{align}

We can then consider the system of IBPs for $d=2$. It is easy to check that in this
case $3$ MIs degenerate, and therefore only $5$ MIs remain linearly independent. 
We do not
report here the equivalent of relations~\eqref{eq:relban2mA}~\eqref{eq:relban2mB1} and~\eqref{eq:relban2mB2},
since they are considerable more lengthy, but one can easily work out the reduction
in $d=2$
and find that, for example, $\I_6(2;s)$, $\I_7(2;s)$ and $\I_8(2;s)$ can be written
as linear combinations of the $\I_1(2;s)$,...,$\I_5(2;s)$.
Using the methods described above, 
$3$ out of the $8$ differential equations for this particular mass configuration
can be decoupled in the limit $d \to 2$, and one can in principle derive a fifth-order
differential equation for any of the MIs, and in particular for the scalar
amplitude $\I_1(d;s)$. 

\subsubsection{The general case of four different masses}
Last but not least, we move to considering the most general configuration with four different masses.
In this case the complexity increases even further and 
solving the IBPs in $d$ dimensions brings to a reduction in terms of 11 different MIs
\begin{align}
\I_1(d;s) = I_4(d;1,1,1,&1,0,0,0,0,0)\,, \quad  \I_2(d;s) = I_4(d;2,1,1,1,0,0,0,0,0)\,,\nonumber \\
 \I_3(d;s) = I_4(d;1,2,1,&1,0,0,0,0,0)\,,\quad  \I_4(d;s) = I_4(d;1,1,2,1,0,0,0,0,0)\,,\nonumber \\
  \I_5(d;s) = I_4(d;1,1,1,&2,0,0,0,0,0)\,, \quad \I_6(d;s) = I_4(d;3,1,1,1,0,0,0,0,0)\,,\nonumber \\
  \I_7(d;s) = I_4(d;2,2,1,&1,0,0,0,0,0)\,, \quad \I_8(d;s) = I_4(d;2,1,2,1,0,0,0,0,0)\,,\nonumber \\
  \I_9(d;s) = I_4(d;2,1,1,&2,0,0,0,0,0)\,, \quad \I_{10}(d;s) = I_4(d;1,2,2,1,0,0,0,0,0)\,, \nonumber \\
 & \I_{11}(d;s) = I_4(d;1,2,1,2,0,0,0,0,0)\,. 
\end{align}
The number of independent master integrals is obviously very large and, if all differential equations
for the $11$ MIs were to be coupled, this would imply an $11$-th order
differential equation for any of the masters and in particular for the scalar
amplitude $\I_1(d;s)$. It is therefore very interesting in this case to know how many
MIs can be decoupled using the methods described above.
Again it is enough to repeat the reduction to MIs, but fixing this time $d=2$, and we
immediately find that 5 out of the 11
MIs become linearly dependent and can be expressed in terms of the other 6. 
Which MIs survive depends of course on the
internal algorithm for the solution of the IBPs. In our case we find as independent MIs
\begin{align}
\I_1(2;s) = I_4(2;1,1,1,&1,0,0,0,0,0)\,, \quad  \I_2(2;s) = I_4(2;2,1,1,1,0,0,0,0,0)\,,\nonumber \\
 \I_3(2;s) = I_4(2;1,2,1,&1,0,0,0,0,0)\,,\quad  \I_4(2;s) = I_4(2;1,1,2,1,0,0,0,0,0)\,,\nonumber \\
  \I_5(2;s) = I_4(2;1,1,1,&2,0,0,0,0,0)\,, \quad \I_6(2;s) = I_4(2;3,1,1,1,0,0,0,0,0)\,. 
\end{align}
This implies that, the new basis of 11 $d$ dimensional MIs defined following the recipe given
above, fulfils a system of $11$ differential equations, $5$ of which
decouple from the system as $d \to 2$. In this way we expect that
a sixth-order differential equation for the scalar amplitude can be derived in $d=2$.

\subsection{Comments and open questions}
\labbel{sec:Comments}
Before moving on to the conclusions we would like to bring attention to some
issues that might have gone unnoticed and which nevertheless leave room to very interesting
open questions.
In the previous sections we have worked out different applications of the ideas outlined
in Section~\ref{sec:IBPsN}. We have seen explicitly that studying the IBPs
in $d=2$ or $d=4$ can provide, for different Feynman graphs, identities 
useful to decouple the system of differential equations they fulfil. 
However in this discussion there is a point that we have avoided mentioning on purpose.
Let us imagine to have to deal with a Feynman graph with three master integrals
$\I_1$, $\I_2$ and $\I_3$,
which fulfil a system of three coupled differential equations in the limit $d \to 4$,
and let us suppose to apply the methods described above in order to try and
decouple the system. We can imagine that by solving the IBPs in $d=4$ only one 
relation can be found. In this case we know that we can use it in order to decouple
one of the three integrals in the limit $d \to 4$, leaving therefore a system of two coupled
equations, equivalent to a second order differential equation. Is this enough to say that
 there must be no other way to 
fully decouple all three differential equations? The answer is, in general, of course no.
We have discussed already how the decoupling of the differential equations in \textsl{any
even integer} number of dimensions is equivalent to the decoupling of the latter in $d=4$.
In Appendix~\ref{App:Dim} we show explicitly how, if one can find a basis that decouples
the differential equations in $d=2\,n$, a corresponding basis can be constructed which decouples
them in $d=4$. Let us then go back to the problem of the three
coupled master integrals. Let us imagine that studying the IBPs in $d=2$ two linearly independent
relations are found among the three masters and that this allows to completely decouple the system
in $d=2$. In this case we know that a corresponding basis would have to exist in $d=4$ as well,
and we could find it with the methods described in Appendix~\ref{App:Dim}. 
On the other hand, if we had not found any new relations in $d=2$, we could still decide to try in $d=6$,
or in $d=8, 10, 12$ etc. apparently without an end. Who or what tells us when to stop and, therefore, when
we can assert without any doubts that not enough relations can be found, in any even number of dimension,
to decouple completely the system? This question is extremely interesting and we unfortunately
do not have a conclusive answer to it. For sure in all examples considered so far it has always
been enough to study the differential equations in $d=2$ and $d=4$ only in order to find
all needed relations to decouple the system. In those cases where not all equations could be 
decoupled (see for example Sections~\ref{sec:sun2},~\ref{sec:crossed},~\ref{sec:ban})
an attempt to consider different numbers of dimensions
would simply produce no new relations at all, suggesting that there is no way to further simplify
the problem, at least in this framework. Of course this does not constitute a mathematical proof
in any respect. 
If these considerations have not brought to a definite answer yet, they
nevertheless open the possibility for a  different perspective in the way a system
of differential equations for master integrals should be studied. We usually tend to
think that the only physically relevant results are obtained when studying the system in the
limit $d \to 4$. A lot of useful information, though, can be extracted studying the
system as $d \to 2\,n$ and, sometimes, the relations found in this way 
appear to be independent and, in a sense, complementary.
Whether these relations are really independent and how to determine the \textsl{maximum
amount} of information that can be extracted by studying the IBPs in fixed integer numbers
of dimensions remain open questions for now. 
It seems however reasonable to think that a more global approach, which allowed to study
the systems of differential equations in general for any even number of dimensions (and not
only in the limit $d \to 4$) could possibly bring a deeper insight in the structure and properties
of the latter.

\section{Conclusions}  \labbel{sec:Concl} \setcounter{equation}{0} 
\numberwithin{equation}{section}
The method of differential equations has proven to be one of the most
effective and promising tools for the evaluation of multi-loop and multi-scale 
Feynman integrals. The usual procedure consists in reducing all Feynman integrals to
a basis of master integrals through integration by parts identities,
then derive differential equations satisfied by the master integrals 
and finally try and solve them as Laurent expansion in $(d-4)$.
For many problems of physical interests
the coefficients of the Laurent expansion of the master integrals can be expressed in terms of 
a particular class of special functions called multiple polylogarithms. 
It has been noted that, whenever this is possible,
a basis of master integrals can be found such that their differential equations
become triangular in the limit $d \to 4$, allowing a simple integration of the differential equations
by quadrature. It was moreover conjectured that in all such cases a canonical basis can be
found, turning the integration of the differential equations into a straightforward 
algebraic problem. If a complete set of boundary conditions is also known, the problem can therefore be considered
as completely solved. 
On the other hand, different cases are known where such a simplification
cannot be achieved and a minimum number of differential
equations remain coupled. Of course in all these cases also a canonical
basis (in the original sense introduced in~\cite{Henn:2013pwa}) cannot be found. 
Whenever this happens, it becomes of crucial importance to be able to determine the \textsl{minimum
number} of master integrals which cannot be decoupled from the system.
If two or more equations are coupled, in fact, no general technique exists to find a solution and one
must resort to different considerations in order to find a complete set of homogeneous solutions, which
can then be used in order to build up the inhomogeneous solution using Euler's method of the variation
of constants (see for example~\cite{Laporta:2004rb}). Of course, the larger the number of coupled
equations is, the more difficult it becomes finding a complete set of solutions. Reducing the
order of the system of differential equations is therefore essential from a practical point of view in order
to be able to successfully tackle the problem.
The issue is nevertheless interesting also from a more general point of view.
Master integrals satisfying higher order differential equations, in fact, usually cannot be expressed in terms
of multiple polylogarithms only and a very intensive theoretical effort has been recently devoted
to determine the properties of the new special functions required. 
The most famous example is the two-loop massive sunrise graph. In this case two differential equations
remain coupled and therefore the problem amounts to solving a second-order differential equation. 
It has been recently shown that the solution of the latter can be expressed in terms
of a new generalisation of the multiple polylogarithms, called elliptic polylogarithms.
Many questions are nevertheless still to be answered. Are elliptic polylogarithms enough for describing
all Feynman integrals whose evaluation can be reduced to a second order differential equation? And what
about higher order equations? 

A first step towards an answer to these questions seems therefore to be in
a criterion to determine, given a set of master integrals and the system of differential equations they fulfil, 
the minimum number of differential equations coupled, and therefore the 
class of special functions required.
In this paper we presented a simple idea which proved to be very useful in this respect.
We showed in particular that the study of the IBPs for fixed integer values of the space-time dimensions, $d=n$, 
can provide the information required for decoupling the differential equations in the limit $d \to n$.
Indeed our criterion is, in principle, a sufficient but not a \textsl{necessary} one, 
in the sense that we did not prove that
if no extra relation can be found among the MIs when $d = n$, then no decoupling is possible for $d \to n$. 
The criterion has moreover proven to be extremely effective, inasmuch as it provided, 
in all cases that we considered, relations useful for decoupling some of the differential equations
and therefore substantially simplify the problem at hand. 
It would indeed be extremely interesting to prove whether this criterion 
is not only a sufficient but also a necessary criterion, checking, for example, whether the number
of independent MIs of the three-loop banana graph in $d=2$ (see Section~\ref{sec:ban}) 
can be further reduced by any other means, reducing in this way also the maximum degree of the differential equation
satisfied by the scalar amplitude.
The criterion is moreover extremely simple to apply, since it can be very easily implemented into any existing 
public or private IBPs reduction code.
In this respect, the possibility of pairing the study of IBPs in fixed numbers
of space-time dimensions together with the new concept of a (pseudo-)finite basis of MIs, 
recently introduced in~\cite{vonManteuffel:2014qoa}, looks particularly promising. 
The latter, in fact, could potentially provide a way to automatically determine the 
highest poles developed by different Feynman integrals for different values of the space-time dimensions.

\section*{Acknowledgements}
I am indebted to Ettore Remiddi, whose continuous advice and support were fundamental for 
the completion of this work. The idea of looking for identities between master integrals for fixed numbers
of the space-time dimensions in order to simplify the systems of differential equations was first developed with
him in~\cite{Remiddi:2013joa}. The generalisation of those ideas presented in this paper 
have benefited from many interesting discussions with him.
I wish to thank Andreas von Manteuffel for the continuous encouragement
to follow the ideas developed in this paper and for having allowed me to use the
development version of Reduze 2 prior to its publication. 
Finally I need to thank Pierpaolo Mastrolia for many discussions and useful
input during different stages of the project, Thomas Gehrmann for his continuous support and for 
his comments on the manuscript and Dominik Kara for carefully proofreading 
the present version of the paper.

\appendix

\section{Comparison with the Schouten Identities}\labbel{App:Sch} \setcounter{equation}{0} 
\numberwithin{equation}{section}
In this appendix we would like to show explicitly how the methods described in this paper
are equivalent to the Schouten Identities
introduced in~\cite{Remiddi:2013joa}.
We will consider again the 
two-loop sunrise with one massive and two massless propagators (see Section~\ref{sec:sun1}
for the definitions of the MIs) and try to derive relation~\eqref{eq:relsun1d2} using the Schouten Identities.
The two-loop sunrise graph depends on three independent momenta, the two loop
momenta $k$, $l$ and the external momentum $p$. With these three momenta we can
build up a $d$-dimensional Schouten polynomial which becomes zero as $d \to n$ with $n \in \mathbb{N}$
and $n \leq 2$.
Following~\cite{Remiddi:2013joa} we start off by considering the quantity
\begin{equation}
\epsilon(k,l,p) = \epsilon_{\mu\nu\rho}\,k^\mu l^\nu p^\rho \labbel{eq:gram}
\end{equation}
defined in $d=3$ space-time dimensions. Indeed~\eqref{eq:gram} is nothing but
the Gram determinant of the three vectors $k,l,p$. By squaring~\eqref{eq:gram} in $d=3$
we obtain a polynomial
\begin{align}
P_2(d;k,l,p) &= ( \epsilon_{\mu \nu \rho}k^\mu l^\nu p^\rho )^2 \nonumber \\ &=
k^2\, l^2 \, p^2 - k^2 \, (l \cdot p)^2 
- p^2 \,(k \cdot l)^2  -  l^2\,(k \cdot p)^2 +  2(k \cdot l)\,( k \cdot p) \, (l \cdot p)\,.
\end{align}
The polynomial was obtained in $d=3$ dimensions, 
but since it contains only scalar products of the three momenta it can be 
easily analytically continued to $d$ dimensions and regarded as a $d$ dimensional
polynomial. By construction the polynomial is zero as $d \to 2$
$$P_2(d \to 2;k,l,p) \to 0\,, \quad \mbox{i.e.}\quad P_2(d\to2;k,l,p) = \mathcal{O}(d-2)\,.$$

Let us consider now the two master integrals defined in~\eqref{eq:missun1}. As 
discussed already both masters develop a double pole in $(d-2)$
\begin{align}
\I_j(d;p^2) = \frac{1}{(d-2)^2}\, \I_j^{(-2)}(2;p^2) + \frac{1}{(d-2)}\,\I_j^{(-1)}(2;p^2) + \I_j^{(0)}(2;p^2) + ...\,,
\quad \mbox{with}\; j=1,2\,.
\end{align}
We consider now the following quantities
\begin{equation}
Z(d;n_1,n_2,n_3) = \int \D^d k  \D^d l\, \frac{P_2(d;k,l,p) }
{\left( k^2 \right)^{n_1} \left( l^2 \right )^{n_2} \left((k-l+p)^2-m^2 \right)^{n_3}}\,,
\end{equation}
which are of course linear combinations of integrals belonging to the sunrise graph, equation~\eqref{eq:sunr1}.
The Schouten polynomial goes to zero as $d \to 2$
and provides therefore an additional factor $(d-2)$ in the numerator, 
which can be used in order to alleviate the total divergence of the integral. 
Assume now that, using this piece of information, 
we can prove that the integral $Z(d;n_1,n_2,n_3)$, for a given choice of
the indices $\{n_1, n_2, n_3\}$, can develop at most a single pole in $(d-2)$
\begin{align}
Z(d\to2;n_1,n_2,n_3) \propto \mathcal{O}\left( \frac{1}{d-2} \right)\,.
\end{align}

If this is true then, for any values of the indices $\{n_1,n_2,n_3\}$, we can consider the integral $Z(d;n_1,n_2,n_3)$
as an integral of the sunrise family and use $d$-dimensional IBPs to reduce it to the two MIs
\begin{equation}
Z(d;n_1,n_2,n_3) = C_1^{n_1n_2n_3}(d;p^2)\,\I_1(d;p^2) 
+ C_2^{n_1n_2n_3}(d;p^2)\,\I_2(d;p^2)\,, \labbel{eq:redsch}
\end{equation}
where the $C_j^{n_1n_2n_3}(d;p^2)$ are in general rational functions of $p^2$ and $d$.
Suppose now that the $C_j^{n_1n_2n_3}(d;p^2)$ do not develop any \textsl{overall} factor $(d-2)$ in the numerator.
If this is the case, upon expanding in Laurent series both the left- and right-hand side of~\eqref{eq:redsch} and using
the fact that the l.h.s. has only a single pole, we can find a relation between the double poles of the
two MIs $\I_1$ and $\I_2$.
To see how this works in practice, let us study this example for some specific choices 
of the indices $n_1,n_2,n_3$ and see how this piece of information can be easily read out.

\begin{itemize}
\item[a)] 
We start considering the easiest case $n_1=n_2=n_3 = 1$, i.e. we study the integral
$$Z(d;1,1,1)\int \D^d k  \D^d l\, \frac{P_2(d;k,l,p) }
{k^2  \, l^2 \, \left((k-l+p)^2-m^2 \right)}\,.$$

Since $I(d;1,1,1,0,0)$ has a maximum pole $1/(d-2)^2$,
we could naively expect that, due to the overall $(d-2)$ factor carried by the
Schouten polynomials, $Z(d;1,1,1)$ should develop at most a single pole $1/(d-2)$.
This is in general of course not granted, since the polynomial in the numerator can
worsen the UV behaviour of the integral.
This naive expectation can be nevertheless easily verified by different means, for example using
sector decomposition~\cite{Borowka:2015mxa}.
On the other hand, as already
specified above, the two MIs have both a double pole $1/(d-2)^2$. 
Expressing $Z(d;1,1,1)$ in terms of $\I_1$ and $\I_2$ one finds easily that the
corresponding coefficients $C_1^{111}(d;p^2)$, $C_2^{111}(d;p^2)$ do not have any overall
$(d-2)$ factor. Therefore we can
expand both the left- and the right-hand side in Laurent series in
$(d-2)$ and keeping only the first orders we get
\begin{align}
\mathcal{O}\left( \frac{1}{d-2} \right) =
\frac{1}{(d-2)^2}\, \frac{(m^2)^2\,p^2}{6}\, 
\left( \I_1^{(-2)}(2;p^2) - (p^2 - m^2) \I_2^{(-2)}(2;p^2) \right) 
+ \mathcal{O}\left( \frac{1}{d-2} \right)\,. \labbel{Z111}
\end{align}
Eq.~\eqref{Z111} gives for consistency a relation between the double poles of the
two MIs, i.e.
$$\I_1^{(-2)}(2;p^2) - (p^2 - m^2) \I_2^{(-2)}(2;p^2) = 0\,,$$
which is, as expected, identical to the relation obtained studying the IBPs in $d=2$, Eq.~\eqref{eq:relsun1d2}.

\item[b)] As a second example, let us consider the case
$n_1 = n_3 = 2\,, \quad n_2=1$  
(or, equivalently, $n_2= n_3 = 2$\,, $n_1=1$ ).
Also in this case $I(d;2,1,2,0,0)$ has a double pole in $(d-2)$
and we would naively expect that $Z(d;2,1,2)$ should therefore develop again at most a single pole. 
This naive expectation can be once more verified explicitly using, for example, sector decomposition.
Expressing $Z(d;2,1,2)$ as linear combination of MIs and expanding in $(d-2)$ we get
\begin{align}
\mathcal{O}\left( \frac{1}{d-2} \right) = -
\frac{1}{(d-2)^2}\, \frac{m^2}{12}\, 
\left( \I_1^{(-2)}(p^2) - (p^2 - m^2) \I_2^{(-2)}(p^2) \right) 
+ \mathcal{O}\left( \frac{1}{d-2} \right)\,, \labbel{Z212}
\end{align}
which for consistency implies
$$\I_1^{(-2)}(2;p^2) - (p^2 - m^2) \I_2^{(-2)}(2;p^2) = 0\,,$$
in agreement with the previous case.

\item[c)]  
As last example we can check what happens for the combination of indices $n_1 = n_2 = 2\,, \quad n_3=1$.
Again, repeating all considerations above, one finds that $I(d;2,2,1,0,0)$ 
has a double pole $1/(d-2)^2$ and  
 $Z(d;2,2,1)$ has instead only a single pole $1/(d-2)$. Reducing $Z(d;2,2,1)$ and expanding in $(d-2)$
 we find
 \begin{align}
\mathcal{O}\left( \frac{1}{d-2} \right) =
\frac{1}{(d-2)^2}\, \frac{m^2}{6}\, 
\left( I_1^{(-2)}(p^2) - (p^2 - m^2) I_2^{(-2)}(p^2) \right) 
+ \mathcal{O}\left( \frac{1}{d-2} \right)\,, \labbel{Z221}
\end{align}
which once more implies
$$\I_1^{(-2)}(2;p^2) - (p^2 - m^2) \I_2^{(-2)}(2;p^2) = 0\,,$$
again in agreement with our previous findings.
\end{itemize}

Using the Schouten Identities one recovers therefore the same relation found by studying the IBPs
in the limit $d=2$. The relation can then be used, as shown in Section~\ref{sec:Tri2}, in order to 
decouple the system of differential equations for this two-loop sunrise graph.
We want to stress here that, differently from the direct study of the IBPs in $d=n$, Schouten 
identities can be derived only if a sufficient number of independent momenta exist.
In the case of the two-loop sunrise studied here there are three independent momenta and we can therefore
derive a Schouten identity in $d=2$ only, but we have no mean to study possible relations
among the MIs in $d=4$. On the other hand, by studying the IBPs in fixed number of
dimensions, one can easily try and look for relations in \textsl{any} number of dimensions,
and in particular in $d=4$, see Section~\ref{sec:sun1}.

\section{Dimensional shift of systems of differential equations} \labbel{App:Dim} \setcounter{equation}{0} 
\numberwithin{equation}{section}
In this paper we showed that by studying the IBPs for fixed numbers of space-time dimensions $d=n$,
one can in general find relations which allow to decouple some of the master integrals from the system
of differential equations, and simplifying therefore the solution of the latter.
In physical applications we are of course interested in the case $d = 4$.
On the other hand we have already briefly discussed how, once the full set of MIs is known 
as Laurent expansion for $d \to 2\,n$, with $n \in \mathbb{N}$, by using Tarasov-Lee
shift identities~\cite{Tarasov:1996br,Lee:2009dh} one can reconstruct 
their Laurent expansion in any other even number of
dimensions, and in particular in $d = 4$. 
It is therefore clear that, if by any means one can find a basis of MIs whose differential equations
are in a \textsl{convenient} form as $d \to 2\,n$ (triangular form, canonical form, etc.), then there must exist
a corresponding basis of MIs whose differential equations look \textsl{identical}
under the formal substitution $(d-2\,n) \to (d-4)$. In this appendix we want to show how this
is indeed true and that such a  basis can be obtained straightforwardly 
by a repeated use of Tarasov-Lee identities.

Let us start considering a topology with $N$ MIs $M_j(d;x_{ij})$ where $j=1,...,N$ and
we made explicit the dependence on the dimensions $d$ and on the
invariants of the problem $x_{ij} = p_i\cdot p_j$. 
Let us assume that, similarly to the case of the two-loop massive sunrise, Section~\ref{sec:sun2},
we are able to find a basis of MIs such that the differential equations take a particularly convenient
form in the limit $d \to 2$. In particular, in order to simplify the notation, let us assume that the differential
equations are linear in $d$ and can be written as

\begin{align}
\frac{\partial}{\partial x_{ij}} M_j(d;x_{ij}) &= A_{0}(2;x_{ij}) \, M_j(d;x_{ij})
+ (d-2)\, A_{1}(2;x_{ij})\, M_j(d;x_{ij}) \,, \labbel{eq:sys1}
\end{align}
where $A_0(2;x_{ij})$ is an $N \times N$ triangular matrix, which
does not depend on $d$, while $A_1(2;x_{ij})$ does not need to be triangular.
If the system of differential equations is in this form, then by solving the
homogeneous system (whose solution is now easier since the matrix $A_0$ is triangular)
\begin{align}
\frac{\partial}{\partial x_n} H_j(d;x_{ij}) &= A_{0}(2;x_{ij}) \, H_j(d;x_{ij})\,,
\end{align}
we can find a  transformation that puts the system of equations in the form
\begin{align}
\frac{\partial}{\partial x_n} M'_j(d;x_{ij}) &= (d-2)\,B(x_{ij}) \, M'_j(d;x_{ij})\,, \labbel{eq:syscan1}
\end{align}
where $B_n(x_{ij})$ is an $N \times N$ matrix whose entries do not depend on $d$.
Note that this does not ensure \textsl{per se} that the system will be in canonical form, since the
entries of the matrix $B_n(x_{ij})$ might not be in d-log form, but could contain more complicated
functions of the external invariants. This form is nevertheless very convenient
for the explicit integration of the equations as Laurent series in $(d-2)$.

In physical applications we are usually interested in the master
integrals expanded in $(d-4)$, which are the physical space-time dimensions.
Indeed we might think of solving the system~\eqref{eq:syscan1} as Laurent
expansion in $(d -2)$, and then transport back the results to $(d-4)$ using
Tarasov-Lee identities.
Nevertheless we might also try and proceed differently and use Tarasov-Lee identities
directly at the level of the differential equations in order to determine a new basis of MIs
whose differential equations are identical to~\eqref{eq:syscan1} with the formal replacement 
$(d-2) \to (d-4)$.
Tarasov  shifting relations indeed contain this piece of information. 
By applying Tarasov  shifting operators on the $N$ MIs $M_j(d,x_{ij}) $ 
and reducing the result to the same set of MIs, we find $N$ relations for the $N$ masters which read
\begin{align}
M'_j(d-2,x_{ij}) = \sum_{l} C_l^{(j)}(d;x_{ij})\, M'_l(d;x_{ij})\, \qquad j=1,...,N\,,
\end{align}
where the $C_l^{(j)}(d,k_x)$ are rational functions of the dimensions $d$
and of the invariants $x_{ij}$.
Define now a new set of $N$ MIs $\I_j(d;x_{ij})$ as
\begin{align}
\I_j(d;x_{ij}) = \sum_{l} C_l^{(j)}(d;x_{ij})\, M'_l(d;x_{ij})\,, \qquad j=1,...,N\,. \label{eq:newMIs}
\end{align}
Because of \eqref{eq:syscan1}, sending $d \to d-2$, we find that
\begin{align}
\frac{\partial}{\partial x_{ij}} M'_j(d-2;x_{ij}) &= 
\frac{\partial}{\partial x_{ij}}\,\left( \sum_{l} C_l^{(j)}(d;x_{ij})\, M'_l(d;x_{ij}) \right)
= (d-4)\,B_n(x_{ij}) \, M'_j(d-2,x_{ij})\,,
\end{align}
which can rephrased in terms of the new MIs as
\begin{align}
\frac{\partial}{\partial x_{ij}} \I_j(d;x_{ij}) &= (d-4)\,B(x_{ij}) \, \I_j(d;x_{ij}). \label{eq:syscan2}
\end{align}
Eq.~\eqref{eq:syscan2} is exactly what we were looking for, namely a system
formally identical to~\eqref{eq:syscan1} under the replacement $(d-2) \to (d-4)$.
While we showed this for a set of differential equations in a very special form~\eqref{eq:syscan1},
the considerations explained above are of course valid for any system of differential equations.
Given a set of masters $\vec{M} = (M_1,...,M_N)$ and their system of differential equations
in matrix form
\begin{align}
\frac{\partial}{\partial x_{ij}} \vec{M}(d;x_{ij}) &= A(d;x_{ij}) \, \vec{M}(d;x_{ij})\,,
\end{align}
where no constraint is applied on the matrix $A(d;x_{ij})$,
by using Tarasov  relations we can define a new basis
\begin{align}
\vec{ \I}(d;x_{ij}) = \vec{M}(d-2;x_{ij}) \,,
\end{align}
which, by construction, fulfils a system of differential equations in the form
\begin{align}
\frac{\partial}{\partial x_{ij}} \vec{\I}(d;x_{ij}) &= A(d-2;x_{ij}) \, \vec{\I}(d;x_{ij})\,.
\end{align}

Of course, using Lee identities (or inverting Tarasov  identities above) one can work
in the opposite direction, shifting $d \to d+2$. Defining the new basis of MIs as
\begin{align}
\vec{\J}(d;x_{ij}) = \vec{M}(d+2,x_{ij})\,,  \qquad j=1,...,N\,, \label{eq:newMIs2}
\end{align}
and following the same argument we find immediately
\begin{align}
\frac{\partial}{\partial x_n} \vec{\J}(d;x_{ij}) &= A(d+2;x_{ij})\, \vec{\J}(d;x_{ij})\,.\label{eq:syscan3}
\end{align}

\newpage 
\bibliographystyle{bibliostyle}   
\bibliography{Biblio}

\providecommand{\href}[2]{#2}\begingroup\raggedright\begin{thebibliography}{10}

\bibitem{'tHooft:1972fi}
G.~'t~Hooft and M.~Veltman, {\it {Regularization and Renormalization of Gauge
  Fields}},  {\em Nucl.Phys.} {\bf B44} (1972) 189--213.

\bibitem{Cicuta:1972jf}
G.~Cicuta and E.~Montaldi, {\it {Analytic renormalization via continuous space
  dimension}},  {\em Lett.Nuovo Cim.} {\bf 4} (1972) 329--332.

\bibitem{Bollini:1972ui}
C.~Bollini and J.~Giambiagi, {\it {Dimensional Renormalization: The Number of
  Dimensions as a Regularizing Parameter}},  {\em Nuovo Cim.} {\bf B12} (1972)
  20--25.

\bibitem{Tkachov:1981wb}
F.~Tkachov, {\it {A Theorem on Analytical Calculability of Four Loop
  Renormalization Group Functions}},  {\em Phys.Lett.} {\bf B100} (1981)
  65--68.

\bibitem{Chetyrkin:1981qh}
K.~Chetyrkin and F.~Tkachov, {\it {Integration by Parts: The Algorithm to
  Calculate beta Functions in 4 Loops}},  {\em Nucl.Phys.} {\bf B192} (1981)
  159--204.

\bibitem{Anastasiou:2004vj}
C.~Anastasiou and A.~Lazopoulos, {\it {Automatic integral reduction for higher
  order perturbative calculations}},  {\em JHEP} {\bf 0407} (2004) 046,
  [\href{http://arxiv.org/abs/hep-ph/0404258}{{\tt hep-ph/0404258}}].

\bibitem{Smirnov:2008iw}
A.~Smirnov, {\it {Algorithm FIRE -- Feynman Integral REduction}},  {\em JHEP}
  {\bf 0810} (2008) 107, [\href{http://arxiv.org/abs/0807.3243}{{\tt
  arXiv:0807.3243}}].

\bibitem{Studerus:2009ye}
C.~Studerus, {\it {Reduze-Feynman Integral Reduction in C++}},  {\em
  Comput.Phys.Commun.} {\bf 181} (2010) 1293--1300,
  [\href{http://arxiv.org/abs/0912.2546}{{\tt arXiv:0912.2546}}].

\bibitem{vonManteuffel:2012np}
A.~von Manteuffel and C.~Studerus, {\it {Reduze 2 - Distributed Feynman
  Integral Reduction}},  \href{http://arxiv.org/abs/1201.4330}{{\tt
  arXiv:1201.4330}}.

\bibitem{Laporta:1996mq}
S.~Laporta and E.~Remiddi, {\it {The Analytical value of the electron (g-2) at
  order alpha**3 in QED}},  {\em Phys.Lett.} {\bf B379} (1996) 283--291,
  [\href{http://arxiv.org/abs/hep-ph/9602417}{{\tt hep-ph/9602417}}].

\bibitem{Laporta:2001dd}
S.~Laporta, {\it {High precision calculation of multiloop Feynman integrals by
  difference equations}},  {\em Int.J.Mod.Phys.} {\bf A15} (2000) 5087--5159,
  [\href{http://arxiv.org/abs/hep-ph/0102033}{{\tt hep-ph/0102033}}].

\bibitem{Kotikov:1990kg}
A.~Kotikov, {\it {Differential equations method: New technique for massive
  Feynman diagrams calculation}},  {\em Phys.Lett.} {\bf B254} (1991) 158--164.

\bibitem{Bern:1993kr}
Z.~Bern, L.~J. Dixon, and D.~A. Kosower, {\it {Dimensionally regulated pentagon
  integrals}},  {\em Nucl.Phys.} {\bf B412} (1994) 751--816,
  [\href{http://arxiv.org/abs/hep-ph/9306240}{{\tt hep-ph/9306240}}].

\bibitem{Remiddi:1997ny}
E.~Remiddi, {\it {Differential equations for Feynman graph amplitudes}},  {\em
  Nuovo Cim.} {\bf A110} (1997) 1435--1452,
  [\href{http://arxiv.org/abs/hep-th/9711188}{{\tt hep-th/9711188}}].

\bibitem{Caffo:1998du}
M.~Caffo, H.~Czyz, S.~Laporta, and E.~Remiddi, {\it {The Master differential
  equations for the two loop sunrise selfmass amplitudes}},  {\em Nuovo Cim.}
  {\bf A111} (1998) 365--389, [\href{http://arxiv.org/abs/hep-th/9805118}{{\tt
  hep-th/9805118}}].

\bibitem{Gehrmann:1999as}
T.~Gehrmann and E.~Remiddi, {\it {Differential equations for two loop four
  point functions}},  {\em Nucl.Phys.} {\bf B580} (2000) 485--518,
  [\href{http://arxiv.org/abs/hep-ph/9912329}{{\tt hep-ph/9912329}}].

\bibitem{Argeri:2007up}
M.~Argeri and P.~Mastrolia, {\it {Feynman Diagrams and Differential
  Equations}},  {\em Int. J. Mod. Phys.} {\bf A22} (2007) 4375--4436,
  [\href{http://arxiv.org/abs/0707.4037}{{\tt arXiv:0707.4037}}].

\bibitem{Gehrmann:2000zt}
T.~Gehrmann and E.~Remiddi, {\it {Two loop master integrals for $\gamma^*
  \rightarrow$ 3 jets: The Planar topologies}},  {\em Nucl.Phys.} {\bf B601}
  (2001) 248--286, [\href{http://arxiv.org/abs/hep-ph/0008287}{{\tt
  hep-ph/0008287}}].

\bibitem{Gehrmann:2001ck}
T.~Gehrmann and E.~Remiddi, {\it {Two loop master integrals for $\gamma^*
  \rightarrow$ 3 jets: The Nonplanar topologies}},  {\em Nucl.Phys.} {\bf B601}
  (2001) 287--317, [\href{http://arxiv.org/abs/hep-ph/0101124}{{\tt
  hep-ph/0101124}}].

\bibitem{Goncharov}
A.~B. Goncharov, {\it {Geometry of configurations, polylogarithms, and motivic
  cohomology}},  {\em Adv. Math.} {\bf 114} (1995), no.~2 197--318.

\bibitem{Remiddi:1999ew}
E.~Remiddi and J.~Vermaseren, {\it {Harmonic polylogarithms}},  {\em
  Int.J.Mod.Phys.} {\bf A15} (2000) 725--754,
  [\href{http://arxiv.org/abs/hep-ph/9905237}{{\tt hep-ph/9905237}}].

\bibitem{Gehrmann:2001pz}
T.~Gehrmann and E.~Remiddi, {\it {Numerical evaluation of harmonic
  polylogarithms}},  {\em Comput.Phys.Commun.} {\bf 141} (2001) 296--312,
  [\href{http://arxiv.org/abs/hep-ph/0107173}{{\tt hep-ph/0107173}}].

\bibitem{Gehrmann:2001jv}
T.~Gehrmann and E.~Remiddi, {\it {Numerical evaluation of two-dimensional
  harmonic polylogarithms}},  {\em Comput.Phys.Commun.} {\bf 144} (2002)
  200--223, [\href{http://arxiv.org/abs/hep-ph/0111255}{{\tt hep-ph/0111255}}].

\bibitem{Vollinga:2004sn}
J.~Vollinga and S.~Weinzierl, {\it {Numerical evaluation of multiple
  polylogarithms}},  {\em Comput.Phys.Commun.} {\bf 167} (2005) 177,
  [\href{http://arxiv.org/abs/hep-ph/0410259}{{\tt hep-ph/0410259}}].

\bibitem{Duhr:2011zq}
C.~Duhr, H.~Gangl, and J.~R. Rhodes, {\it {From polygons and symbols to
  polylogarithmic functions}},  {\em JHEP} {\bf 1210} (2012) 075,
  [\href{http://arxiv.org/abs/1110.0458}{{\tt arXiv:1110.0458}}].

\bibitem{Duhr:2012fh}
C.~Duhr, {\it {Hopf algebras, coproducts and symbols: an application to Higgs
  boson amplitudes}},  {\em JHEP} {\bf 1208} (2012) 043,
  [\href{http://arxiv.org/abs/1203.0454}{{\tt arXiv:1203.0454}}].

\bibitem{Panzer:2014caa}
E.~Panzer, {\it {Algorithms for the symbolic integration of hyperlogarithms
  with applications to Feynman integrals}},  {\em Comput.Phys.Commun.} {\bf
  188} (2014) 148--166, [\href{http://arxiv.org/abs/1403.3385}{{\tt
  arXiv:1403.3385}}].

\bibitem{Kotikov:2010gf}
A.~Kotikov, {\it {The Property of maximal transcendentality in the N=4
  Supersymmetric Yang-Mills}},  \href{http://arxiv.org/abs/1005.5029}{{\tt
  arXiv:1005.5029}}.

\bibitem{Henn:2013pwa}
J.~M. Henn, {\it {Multiloop integrals in dimensional regularization made
  simple}},  {\em Phys.Rev.Lett.} {\bf 110} (2013) 251601,
  [\href{http://arxiv.org/abs/1304.1806}{{\tt arXiv:1304.1806}}].

\bibitem{Caron-Huot:2014lda}
S.~Caron-Huot and J.~M. Henn, {\it {Iterative structure of finite loop
  integrals}},  {\em JHEP} {\bf 06} (2014) 114,
  [\href{http://arxiv.org/abs/1404.2922}{{\tt arXiv:1404.2922}}].

\bibitem{Bern:2014kca}
Z.~Bern, E.~Herrmann, S.~Litsey, J.~Stankowicz, and J.~Trnka, {\it {Logarithmic
  Singularities and Maximally Supersymmetric Amplitudes}},  {\em JHEP} {\bf 06}
  (2015) 202, [\href{http://arxiv.org/abs/1412.8584}{{\tt arXiv:1412.8584}}].

\bibitem{Argeri:2014qva}
M.~Argeri, S.~Di~Vita, P.~Mastrolia, E.~Mirabella, J.~Schlenk, U.~Schubert, and
  L.~Tancredi, {\it {Magnus and Dyson Series for Master Integrals}},  {\em
  JHEP} {\bf 1403} (2014) 082, [\href{http://arxiv.org/abs/1401.2979}{{\tt
  arXiv:1401.2979}}].

\bibitem{DiVita:2014pza}
S.~Di~Vita, P.~Mastrolia, U.~Schubert, and V.~Yundin, {\it {Three-loop master
  integrals for ladder-box diagrams with one massive leg}},  {\em JHEP} {\bf
  09} (2014) 148, [\href{http://arxiv.org/abs/1408.3107}{{\tt
  arXiv:1408.3107}}].

\bibitem{Moser}
J.~Moser, {\em {The order of a singularity in Fuchs' theory}}, vol.~72.
\newblock Mathematische Zeitschrift, 1960.

\bibitem{Lee:2014ioa}
R.~N. Lee, {\it {Reducing differential equations for multiloop master
  integrals}},  {\em JHEP} {\bf 1504} (2015) 108,
  [\href{http://arxiv.org/abs/1411.0911}{{\tt arXiv:1411.0911}}].

\bibitem{Henn:2014qga}
J.~M. Henn, {\it {Lectures on differential equations for Feynman integrals}},
  {\em J.Phys.} {\bf A48} (2015), no.~15 153001,
  [\href{http://arxiv.org/abs/1412.2296}{{\tt arXiv:1412.2296}}].

\bibitem{Hoschele:2014qsa}
M.~Hoschele, J.~Hoff, and T.~Ueda, {\it {Adequate bases of phase space master
  integrals for gg $\to$ h at NNLO and beyond}},  {\em JHEP} {\bf 09} (2014)
  116, [\href{http://arxiv.org/abs/1407.4049}{{\tt arXiv:1407.4049}}].

\bibitem{Gehrmann:2014bfa}
T.~Gehrmann, A.~von Manteuffel, L.~Tancredi, and E.~Weihs, {\it {The two-loop
  master integrals for $q\bar{q} \to VV$}},  {\em JHEP} {\bf 1406} (2014) 032,
  [\href{http://arxiv.org/abs/1404.4853}{{\tt arXiv:1404.4853}}].

\bibitem{Laporta:2004rb}
S.~Laporta and E.~Remiddi, {\it {Analytic treatment of the two loop equal mass
  sunrise graph}},  {\em Nucl.Phys.} {\bf B704} (2005) 349--386,
  [\href{http://arxiv.org/abs/hep-ph/0406160}{{\tt hep-ph/0406160}}].

\bibitem{Adams:2013nia}
L.~Adams, C.~Bogner, and S.~Weinzierl, {\it {The two-loop sunrise graph with
  arbitrary masses}},  {\em J.Math.Phys.} {\bf 54} (2013) 052303,
  [\href{http://arxiv.org/abs/1302.7004}{{\tt arXiv:1302.7004}}].

\bibitem{Remiddi:2013joa}
E.~Remiddi and L.~Tancredi, {\it {Schouten identities for Feynman graph
  amplitudes; The Master Integrals for the two-loop massive sunrise graph}},
  {\em Nucl.Phys.} {\bf B880} (2014) 343--377,
  [\href{http://arxiv.org/abs/1311.3342}{{\tt arXiv:1311.3342}}].

\bibitem{Adams:2014vja}
L.~Adams, C.~Bogner, and S.~Weinzierl, {\it {The two-loop sunrise graph in two
  space-time dimensions with arbitrary masses in terms of elliptic
  dilogarithms}},  {\em J.Math.Phys.} {\bf 55} (2014), no.~10 102301,
  [\href{http://arxiv.org/abs/1405.5640}{{\tt arXiv:1405.5640}}].

\bibitem{Tarasov:1996br}
O.~V. Tarasov, {\it {Connection between Feynman integrals having different
  values of the space-time dimension}},  {\em Phys. Rev.} {\bf D54} (1996)
  6479--6490, [\href{http://arxiv.org/abs/hep-th/9606018}{{\tt
  hep-th/9606018}}].

\bibitem{Lee:2009dh}
R.~Lee, {\it {Space-time dimensionality D as complex variable: Calculating loop
  integrals using dimensional recurrence relation and analytical properties
  with respect to D}},  {\em Nucl.Phys.} {\bf B830} (2010) 474--492,
  [\href{http://arxiv.org/abs/0911.0252}{{\tt arXiv:0911.0252}}].

\bibitem{Georgoudis:2015hca}
A.~Georgoudis and Y.~Zhang, {\it {Two-loop Integral Reduction from Elliptic and
  Hyperelliptic Curves}},  \href{http://arxiv.org/abs/1507.06310}{{\tt
  arXiv:1507.06310}}.

\bibitem{Broedel:2015hia}
J.~Broedel, N.~Matthes, and O.~Schlotterer, {\it {Relations between elliptic
  multiple zeta values and a special derivation algebra}},
  \href{http://arxiv.org/abs/1507.02254}{{\tt arXiv:1507.02254}}.

\bibitem{Huber:2015bva}
T.~Huber and S.~Krankl, {\it {Two-loop master integrals for non-leptonic
  heavy-to-heavy decays}},  {\em JHEP} {\bf 1504} (2015) 140,
  [\href{http://arxiv.org/abs/1503.00735}{{\tt arXiv:1503.00735}}].

\bibitem{Birthwright:2004kk}
T.~Birthwright, E.~W.~N. Glover, and P.~Marquard, {\it {Master integrals for
  massless two-loop vertex diagrams with three offshell legs}},  {\em JHEP}
  {\bf 0409} (2004) 042, [\href{http://arxiv.org/abs/hep-ph/0407343}{{\tt
  hep-ph/0407343}}].

\bibitem{Chavez:2012kn}
F.~Chavez and C.~Duhr, {\it {Three-mass triangle integrals and single-valued
  polylogarithms}},  {\em JHEP} {\bf 1211} (2012) 114,
  [\href{http://arxiv.org/abs/1209.2722}{{\tt arXiv:1209.2722}}].

\bibitem{Gehrmann:2013cxs}
T.~Gehrmann, L.~Tancredi, and E.~Weihs, {\it {Two-loop master integrals for $q
  \bar{q} \to VV$: the planar topologies}},  {\em JHEP} {\bf 1308} (2013) 070,
  [\href{http://arxiv.org/abs/1306.6344}{{\tt arXiv:1306.6344}}].

\bibitem{Henn:2014lfa}
J.~M. Henn, K.~Melnikov, and V.~A. Smirnov, {\it {Two-loop planar master
  integrals for the production of off-shell vector bosons in hadron
  collisions}},  {\em JHEP} {\bf 1405} (2014) 090,
  [\href{http://arxiv.org/abs/1402.7078}{{\tt arXiv:1402.7078}}].

\bibitem{MullerStach:2012mp}
S.~Muller-Stach, S.~Weinzierl, and R.~Zayadeh, {\it {Picard-Fuchs equations for
  Feynman integrals}},  {\em Commun.Math.Phys.} {\bf 326} (2014) 237--249,
  [\href{http://arxiv.org/abs/1212.4389}{{\tt arXiv:1212.4389}}].

\bibitem{Aglietti:2007as}
U.~Aglietti, R.~Bonciani, L.~Grassi, and E.~Remiddi, {\it {The Two loop crossed
  ladder vertex diagram with two massive exchanges}},  {\em Nucl. Phys.} {\bf
  B789} (2008) 45--83, [\href{http://arxiv.org/abs/0705.2616}{{\tt
  arXiv:0705.2616}}].

\bibitem{Borowka:2015mxa}
S.~Borowka, G.~Heinrich, S.~P. Jones, M.~Kerner, J.~Schlenk, and T.~Zirke, {\it
  {SecDec-3.0: numerical evaluation of multi-scale integrals beyond one loop}},
   \href{http://arxiv.org/abs/1502.06595}{{\tt arXiv:1502.06595}}.

\bibitem{Bonciani:2015eua}
R.~Bonciani, V.~Del~Duca, H.~Frellesvig, J.~M. Henn, F.~Moriello, and V.~A.
  Smirnov, {\it {Next-to-leading order QCD corrections to the decay width $H
  \to Z \gamma $}},  {\em JHEP} {\bf 08} (2015) 108,
  [\href{http://arxiv.org/abs/1505.00567}{{\tt arXiv:1505.00567}}].

\bibitem{Gehrmann:2015dua}
T.~Gehrmann, S.~Guns, and D.~Kara, {\it {The rare decay $H\to Z\gamma$ in
  perturbative QCD}},  {\em JHEP} {\bf 09} (2015) 038,
  [\href{http://arxiv.org/abs/1505.00561}{{\tt arXiv:1505.00561}}].

\bibitem{vonManteuffel:2014qoa}
A.~von Manteuffel, E.~Panzer, and R.~M. Schabinger, {\it {A quasi-finite basis
  for multi-loop Feynman integrals}},  {\em JHEP} {\bf 02} (2015) 120,
  [\href{http://arxiv.org/abs/1411.7392}{{\tt arXiv:1411.7392}}].

\end{thebibliography}\endgroup
\end{document}